\documentclass[useAMS,usegraphicx]{mn2e}

\usepackage{multirow}

\title[Simulated spectral states of AGN]{Simulated spectral states of AGN and observational predictions}

\author[Sobolewska et al.]{Ma{\l}gorzata A.
  Sobolewska$^{1}$\thanks{E-mail:msobolewska@cfa.harvard.edu}, Aneta
  Siemiginowska$^1$, Marek Gierli\'nski$^2$\\$^1$Smithsonian
  Astrophysical Observatory, 60 Garden Street, Cambridge, MA 02138,
  USA\\$^2$Department of Physics, University of Durham, South Road,
  Durham DH1 3LE, UK} 

\newcommand{\aj}{AJ} 
\newcommand{\apj}{ApJ}
\newcommand{\apjl}{ApJL} 
\newcommand{\mnras}{MNRAS}
\newcommand{\aap}{A\&A} 

\newcommand{\araa}{ARAA}
 
\newcommand{\nat}{Nature}
\newcommand{\apjs}{ApJS} 
\newcommand{\aspv}{ASP Conf. Ser. Vol.}
\newcommand{\asp}{Astron. Soc. Pac., San Francisco}
\newcommand{\aapr}{A\&AR}

\begin{document}

\bibliographystyle{plainnat}
\date{}

\pagerange{\pageref{firstpage}--\pageref{lastpage}} \pubyear{2010}

\maketitle

\label{firstpage}

\begin{abstract}

  Active galactic nuclei (AGN) and galactic black hole binaries (GBHs)
  represent two classes of accreting black holes. They both contain an
  accretion disc emitting a thermal radiation, and a non-thermal X-ray
  emitting 'corona'. GBHs exhibit state transitions and spectral states are
  characterized by different luminosity levels and shapes of the
  spectral energy distribution (SED). If AGN transitioned in a similar
  way, the characteristic timescales of such transitions would exceed
  $\sim 10^5$ years. Thus a probability to observe an individual AGN
  transiting between different spectral states is very low. In this paper we follow a
  spectral evolution of a GBH GRO~J1655-40 and then apply its SED
  evolution pattern to a simulated population of AGN under a
  reasonable assumption that a large sample of AGN should contain a
  mixture of sources in different spectral states. We model the X-ray
  spectra of GRO~1655-40 with the {\sc eqpair} model and then scale
  the best-fitting models with the black hole mass to simulate the AGN
  spectra. We compare the simulated and observed AGN SEDs to determine
  the spectral states of observed Type 1 AGN, LINER and NLS1
  populations. We conclude that bright Type 1 AGN and NLS1
  galaxies are in a spectral state similar to the soft spectral state
  of GBHs, while the spectral state of LINERs may correspond to the
  hard spectral state of GBHs.  We find that taking into account a
  spread in the black hole masses over several orders of magnitude, as
  in the observed AGN samples, leads to a correlation between the
  X-ray loudness, $\alpha_{\rm ox}$, and a monochromatic luminosity at
  2500\AA. We predict that the $\alpha_{\rm ox}$ correlates positively
  with the Eddington luminosity ratio down to a critical value of
  $\lambda_{\rm crit} = L/L_{\rm E} \approx 0.01$, and that this
  correlation changes its sign for the accretion rates below
  $\lambda_{\rm crit}$.

\end{abstract}



\section{Introduction}
\label{sec:intro}


\begin{figure*}
\centering
\includegraphics[width=4.3cm,angle=-90]{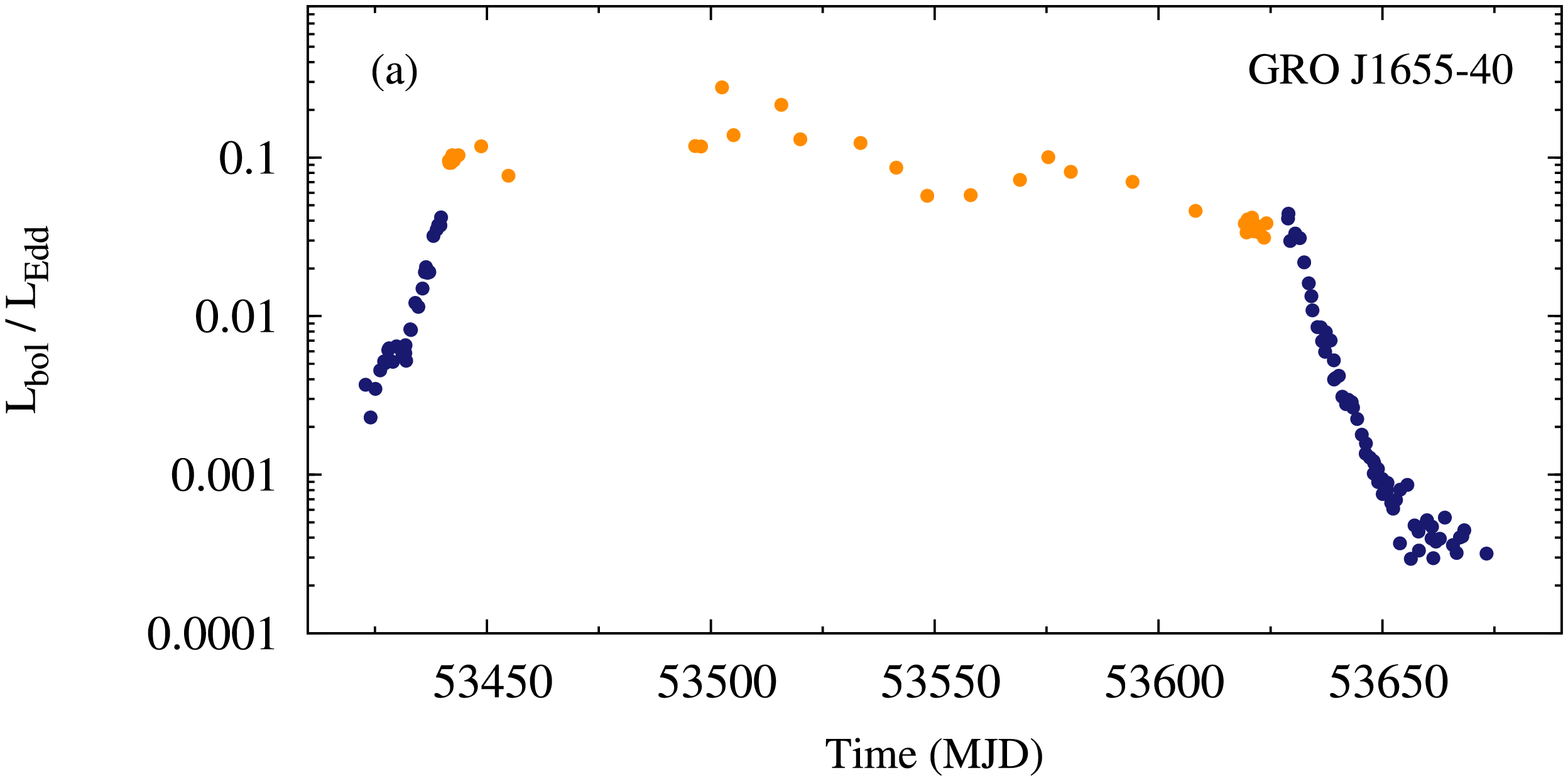}
\includegraphics[width=4.3cm,angle=-90]{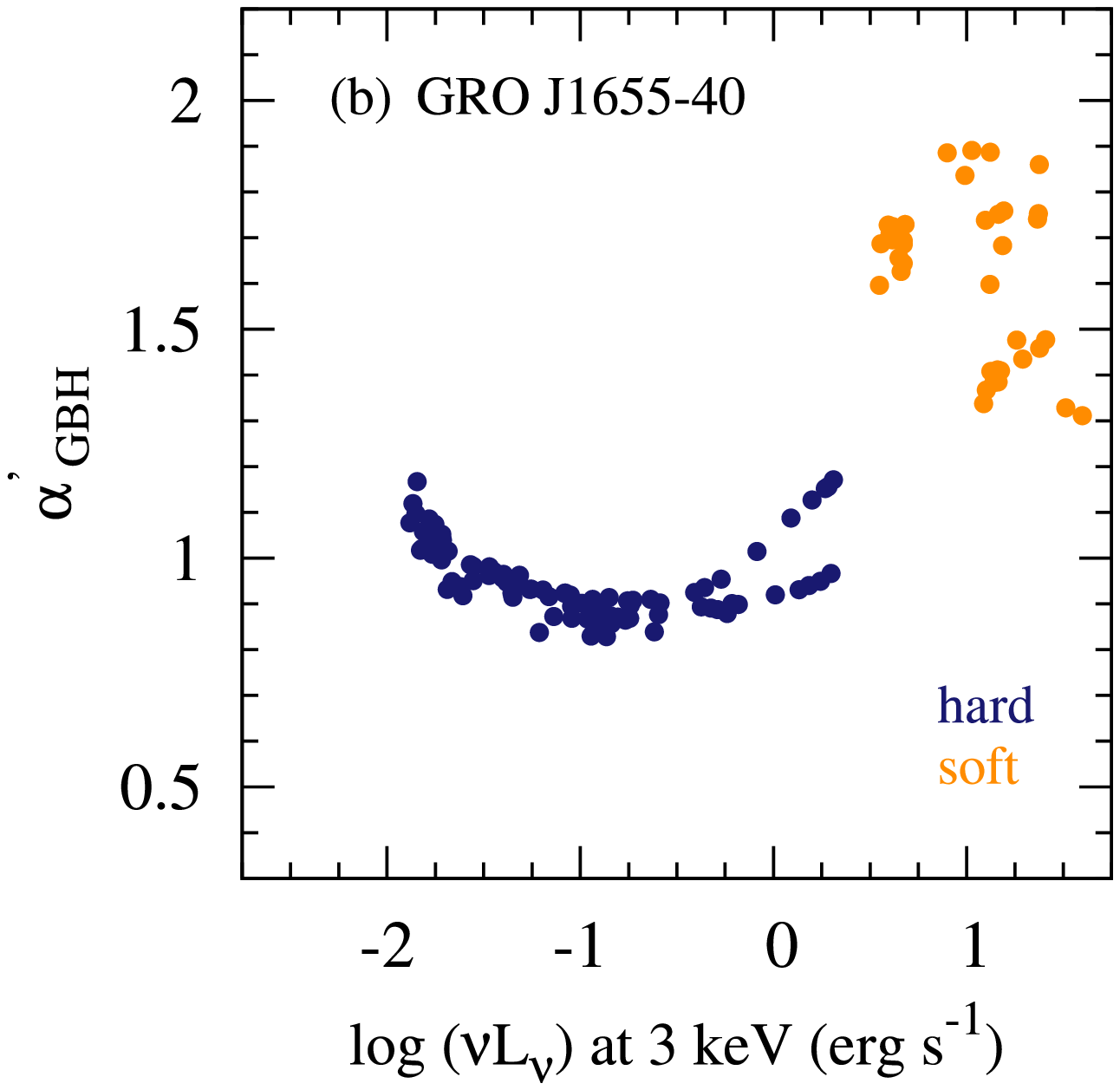}
\includegraphics[width=4.3cm,angle=-90]{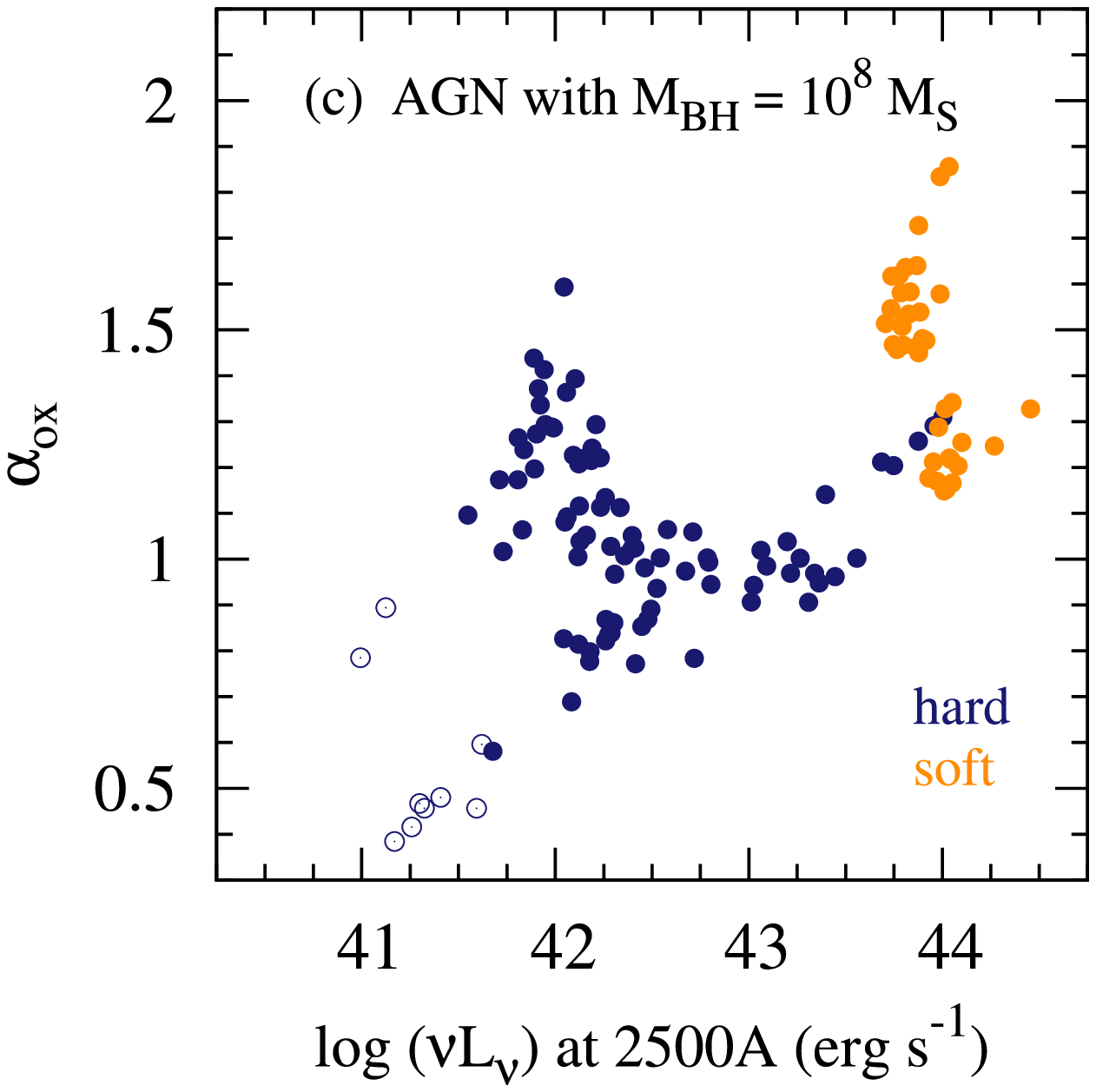}
\caption{Scaling the GRO~J1655-40 outburst to the case of a $10^8$M$_{\odot}$ black
hole in AGN. (a) Lightcurve of GRO J1655-40 with the data used in this paper; hard states
- black/blue, soft state - gray/orange. (b)'Stretched' disc-to-Comptonization index versus the $EF_E$ flux at 3 keV,
$(ef_e)_3$, for GRO~J1655-40 (see Sobolewska, Gierli\'nski, Siemiginowska 2009). (c) Simulated $\alpha_{\rm ox}$ vs.
monochromatic luminosity at 2500\AA. Open circles indicate the results from simulated AGN SEDs based on the GRO~J1655-40
spectra with little or no disc component.}
\label{fig:loop}
\end{figure*}

Active galactic nuclei (AGN) are strong sources of optical and X-ray
radiation. Their emission is commonly thought to be composed of two
main components: thermal blackbody type radiation emerging from a
geometrically thin and optically thick accretion disc in the
optical/UV band, and a non-thermal power-law type hard X-ray tail
believed to be formed by inverse Compton up-scattering of the disc
seed photons by energetic electrons in a hot plasma, so called X-ray
'corona' (e.g. Magdziarz et al. 1998; Chiang 2002; Chiang \& Blaes
2003; Sobolewska, \.Zycki \& Siemiginowska 2004a,b; Middleton, Done \&
Gierli\'nski 2007; Vasudevan \& Fabian 2009).  Dependencies between
the X-ray and optical/UV luminosities, and the shape of the spectral
energy distribution (SED) in different cosmic times are studied to
constrain accretion models, understand AGN variability,  AGN
influence on intergalactic medium and a contribution to the X-ray
background. These dependencies have been investigated by many authors
(e.g. Young, Elvis \& Risaliti 2010; Grupe et al. 2010; Lusso et al.
2010; Zamorani et al. 1981; Avni \& Tananbaum. 1982; Wilkes et al.
1994; Yuan, Siebert \& Brinkmann 1998; Bechtold et al. 2003; Vignali,
Brandt \& Schneider 2003; Strateva et al. 2005; Steffen et al. 2006;
Kelly et al. 2007, 2008). Most of the authors agree that the X-ray
loudness, defined as
\begin{equation}
\alpha_{\rm ox} = - \frac{\log (\nu \rm L_{\nu})_o - \log (\nu \rm L_{\nu})_x}{\log \nu_o - \log \nu_x} + 1,
\end{equation}
where $(\nu L_\nu)_o$ and $(\nu L_\nu)_x$ are monochromatic
luminosities at the rest frame optical/UV and X-ray energies,
$\lambda_o = c/\nu_o = 2500$\AA\/ and $E_x = h\nu_x = 2$ keV,
respectively, correlates primarily with L$_{\nu, o}$, and that there
is a strong correlation between L$_{\nu, o}$ and L$_{\nu, x}$. The
correlation between the rest frame 2--10 keV X-ray photon index,
$\Gamma_x$, and optical/UV or X-ray luminosities is still a matter of
debate, however a correlation of $\Gamma_x$ with the Eddington
luminosity ratio has been reported by several authors (e.g. Grupe
2004, Kelly et al. 2008, Sobolewska \& Papadakis 2009).  Kelly et al.
(2008) pointed out that this correlation may be induced by the
underlying dependence of the X-ray photon index on the mass accretion
rate. Grupe (2006) found a mild correlation between the X-ray loudness
and the X-ray spectral index, $\alpha_x$ = $\Gamma_x - 1$, but its
significance was low. In addition, correlations of $\alpha_{\rm ox}$ and
$\Gamma_x$ with redshift are studied in attempt to discover how
quasars have been evolving over the cosmic time. It seems that
quasars' SEDs become more X-ray quiet for higher redshifts, but no
significant intrinsic dependence of $\Gamma_x$ on the redshift was
found.

Galactic black hole binaries (GBHs) represent another class of
accreting black holes.  Similarly to AGN, they also accrete matter
through an accretion disc emitting thermal X-rays, posses a hard X-ray
emitting 'corona' and may show collimated outflows in the form of
jets, or uncollimated outflows, the so called disc winds. GBHs exhibit
state transitions that can be followed throughout the entire outburst
cycle of the duration of the order of $\sim$100 days. The GBHs state
transitions happen on timescales of hours or shorter. Therefore, X-ray
spectra of GBHs provide a way to study properties of the disc-corona
accretion system at different luminosity levels.

Signatures of AGN outbursts and their different luminosity states have
been recently observed in X-rays (for example observations of clusters
of galaxies, e.g. Fabian et al. 2001, McNamarra \& Nulsen 2007; compact
radio sources, e.g. Czerny et al. 2009; and X-ray jets, e.g.
Siemiginowska et al. 2007). However, in the case of AGN characteristic
transition timescales exceed 10$^5$ years, as the timescales are
thought to scale inversely with the black hole mass, and so AGN
population studies are needed in order to identify sources in
different spectral states (e.g. K{\"o}rding, Jester \& Fender 2006).
One way to deal with this difficulty is to follow GBHs spectral
evolution through different states and then apply its pattern to a
population of AGN to identify their states.

The Rossi X-ray Timing Experiment (RXTE) archives contain GBH data
that cover many complete X-ray outbursts, during which an object emerges
from the quiescence, reaches a significant fraction of Eddington luminosity, and
fades away. Application of even simple models to these data leads to a good
phenomenological understanding of accretion mechanism at different luminosity levels.
Spectral studies accompanied by studies of the high quality X-ray power spectra
facilitate classification of GBHs spectral states 
(e.g. Remillard \& McClintock 2006; Done, Gierli\'nski \& Kubota 2007).

The situation is quite different in the case of AGN. It is commonly
accepted that the typical timescales in accreting black holes are
inversely proportional to the black hole mass, and so the timescales
of state transitions in AGN are too long to be covered by an observing
run. As a consequence, it is unlikely to observe a state transition of
an individual AGN. However, a large enough unbiased sample of AGN
should contain a mixture of objects in different spectral states.
Still, it is not trivial to determine a spectral state of an AGN
relying solely on the X-ray power-law like continuum. The Galactic
black hole binaries phenomenology clearly shows that a SED with a
typical X-ray photon index $\Gamma_x\sim2$ may be representative of a
hard, intermediate, or very high spectral state. Even though these GBH
states show similar hard X-ray spectra, the properties of their
thermal disc components peaking in soft X-rays around 1 keV differ
dramatically (e.g. Remillard \& McClintock 2006; Done et al. 2007).
Since the mass of the accreting black hole affects the characteristic
temperature of the accretion disc emission: the disc spectrum of more
massive objects shifts towards lower energies. As a result AGN's discs
emerge in the optical/UV band. Confirmed GBHs are homogeneous in terms
of the mass of the compact object and they all contain black holes
with mass of the order of several Solar masses (see e.g. Done \&
Gierli\'nski 2003, Remillard \& McClintock 2006, and references
therein). In contrast, the AGN samples contain black holes that differ
in mass by 3--4 orders of magnitude (e.g. Grupe et al. 2010; Merloni
et al. 2010; Shen et al. 2008; Kelly et al. 2008).  Unfortunately,
good quality simultaneous AGN data covering a multiwavelength
optical/UV to hard X-ray band are rarely available. Consequently, it
is not straightforward to disentangle the intrinsic and apparent (e.g.
due to absorption or reflection) AGN X-ray spectral variability and
develop an AGN spectral state phenomenology, similar to that of GBHs.
The method to assign spectral states to AGN based on their power
spectra (see e.g. Uttley \& McHardy 2005, Ar{\'e}valo et al.  2006,
McHardy et al. 2007) can be applied only to a few best monitored
objects with quality light curves. Clearly, there is a need to develop
indirect methods of assigning the accretion mode to the SEDs of AGN.

In Sobolewska, Gierli\'nski, Siemiginowska (2009, hereafter Paper I)
we defined the disc-to-power law index for Galactic black hole
binaries, $\alpha^{ \prime}_{\rm GBH} \simeq 0.32(\alpha_{\rm GBH}
-1)+1$, which corresponds to the X-ray loudness parameter,
$\alpha_{\rm ox}$, used to characterize the AGN broad-band
optical/UV/X-ray spectra. The $\alpha_{\rm GBH} \simeq 0.824
\log[(ef_e)_{\rm 3}/(ef_e)_{\rm 20}] + 1$ (where $ef_e$ denotes the
$EF_E$ flux in erg cm$^{-2}$ s$^{-1}$) was defined as the index of a
nominal power law between the 3 keV, where the accretion disc
dominates the spectrum of a GBH (with the exception of the hard state)
and 20 keV, where the spectrum is dominated by a power law emission,
modelled as Comptonization of the seed disc photons in the hot corona.
The $\alpha^{ \prime}_{\rm GBH}$ accounts for the many orders on
magnitude difference between the energies for which the $\alpha_{\rm
  GBH}$ and $\alpha_{\rm ox}$ are defined (3-to-20 keV and
2500\AA-to-2 keV, respectively). We showed that the distribution of
$\alpha^{ \prime}_{\rm GBH}$ shows three distinctive peaks around
$\sim$1, 1.5 and 2, which correspond to hard, intermediate/very high
and soft spectral states, respectively.  Hence, $\alpha^{\prime}_{\rm
  GBH}$ and so $\alpha_{\rm ox}$, together with the X-ray photon
index, can be indicators of a spectral state of accreting black holes.

In the present paper we develop the Paper I idea in more detail. We
use the GBH phenomenology to simulate an outburst of AGN. We use a
physical model of accretion to describe the spectra of a
representative GBH, GRO~J1655-40, at various luminosity levels, and
then we scale the seed photons temperature and bolometric luminosity
in the best fitting models with mass, as in the standard
Shakura-Sunyaev disc case, to study accretion onto a supermassive
black hole. Next, we separate the simulated AGN spectral states in the
$\alpha_{\rm ox}$ vs. $(\nu L_\nu)_o$ plane for various classes of
AGN, including broad and narrow line Type 1 Seyferts and LINERs. We
investigate if the mass spread in the observed AGN samples can contribute
to the observed correlation between the X-ray loudness, $\alpha_{\rm
  ox}$) and monochromatic luminosity at 2500\AA.

The structure of the paper is as follows. In Sec.~\ref{sec:data} we describe the data
selection criteria and data reduction procedure as well as give details of data
modeling. Section~\ref{sec:method} we present our method of mass scaling of a GBH
outburst to the case of AGN. In Sec.~\ref{sec:res} we present simulated spectra,
resulting correlations between the parameters, and compare our work with observational
samples. Finally, in Sec.~\ref{sec:dc} we discuss and conclude our results.

\begin{figure*}
\centering
\includegraphics[width=4.4cm,angle=-90]{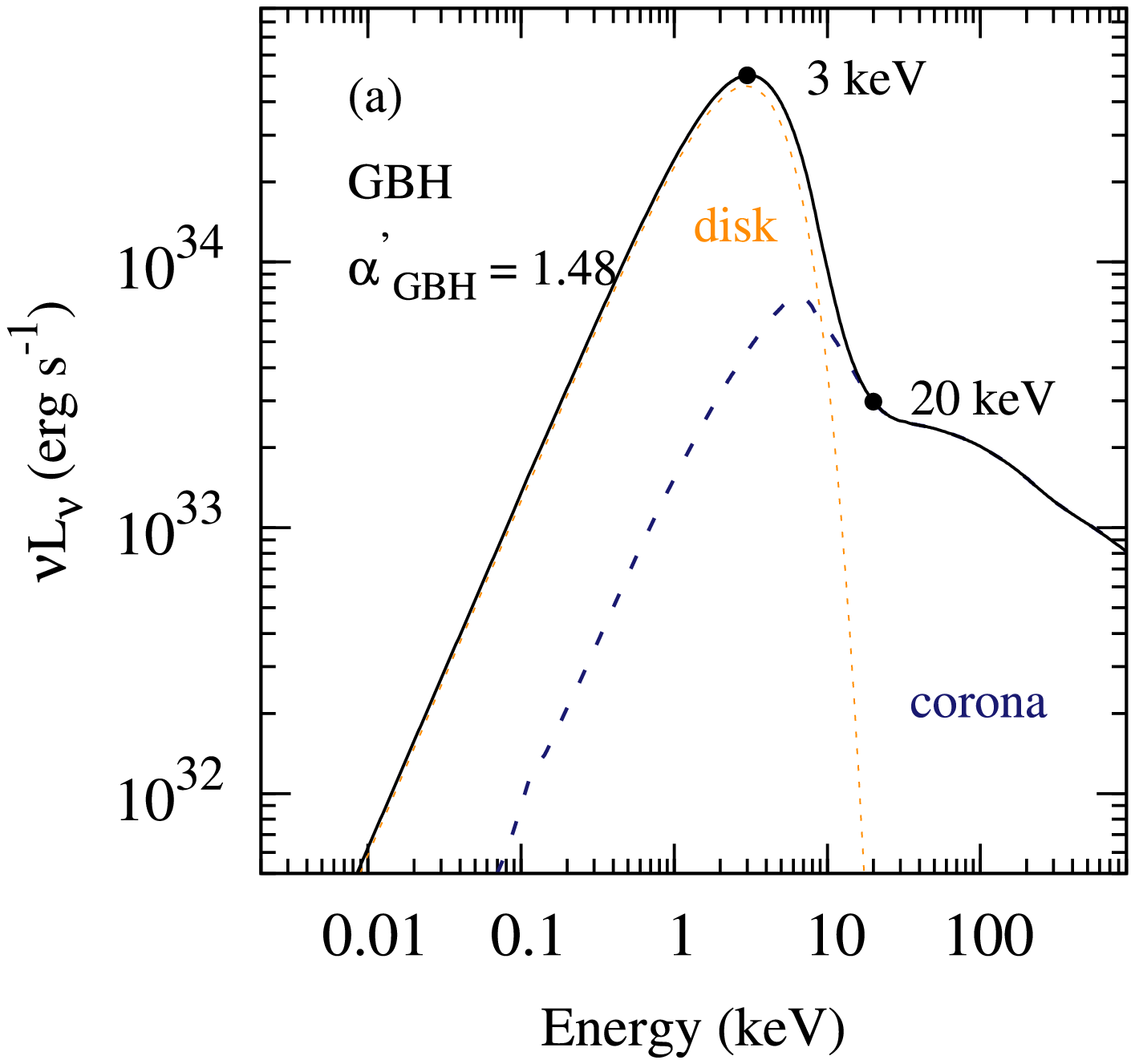}
\includegraphics[width=4.4cm,angle=-90]{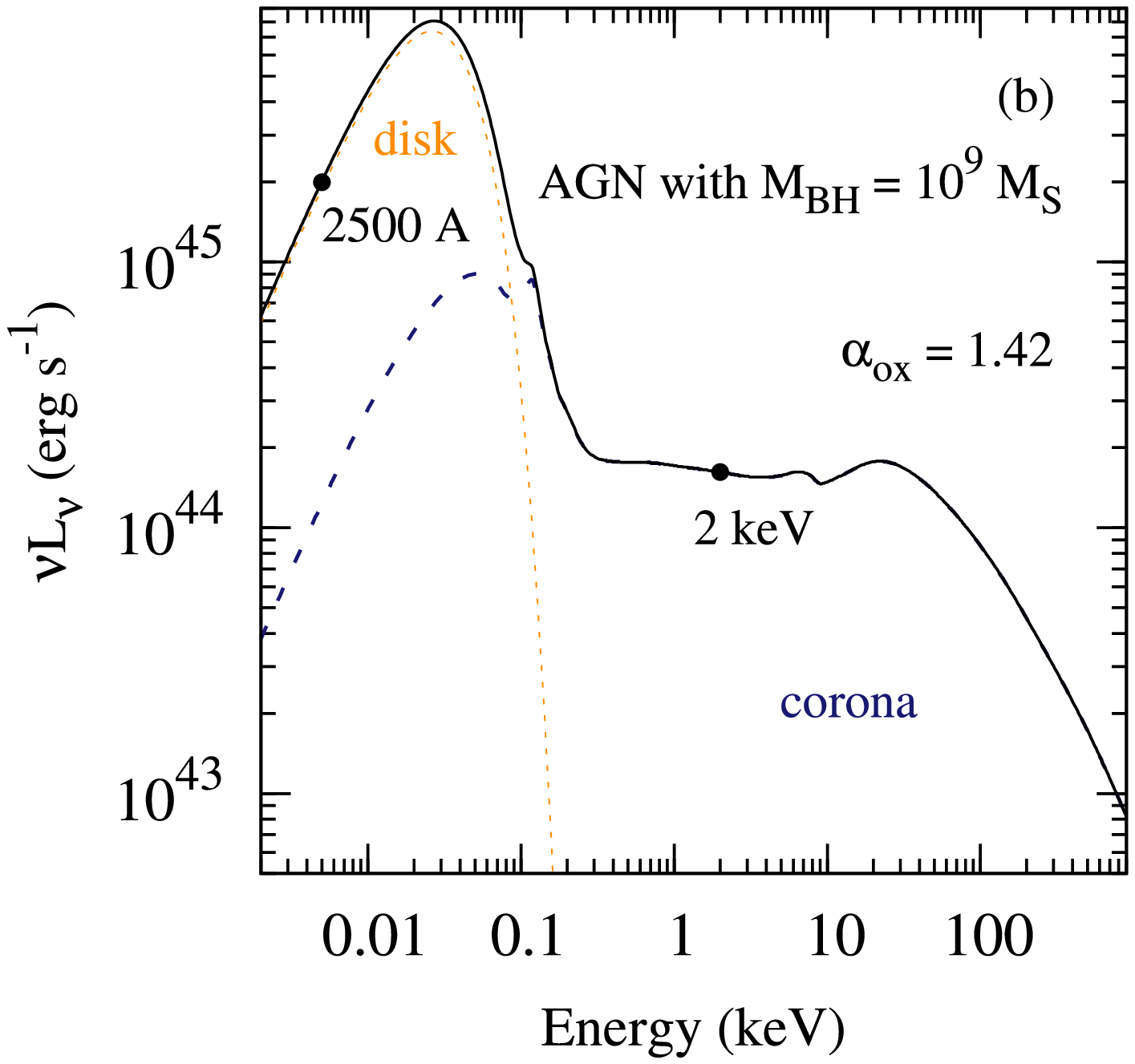}\\
\includegraphics[width=4.4cm,angle=-90]{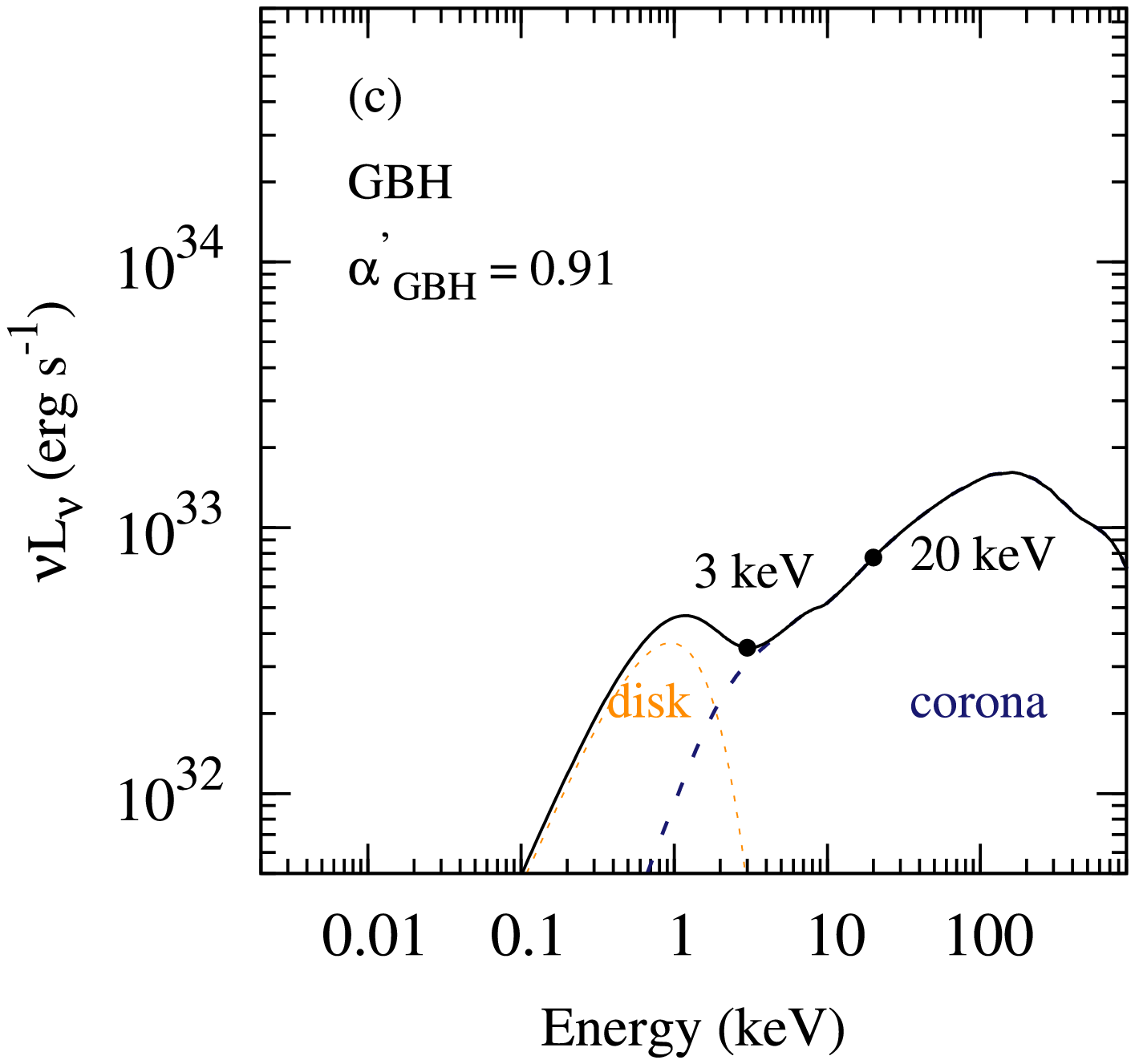}
\includegraphics[width=4.4cm,angle=-90]{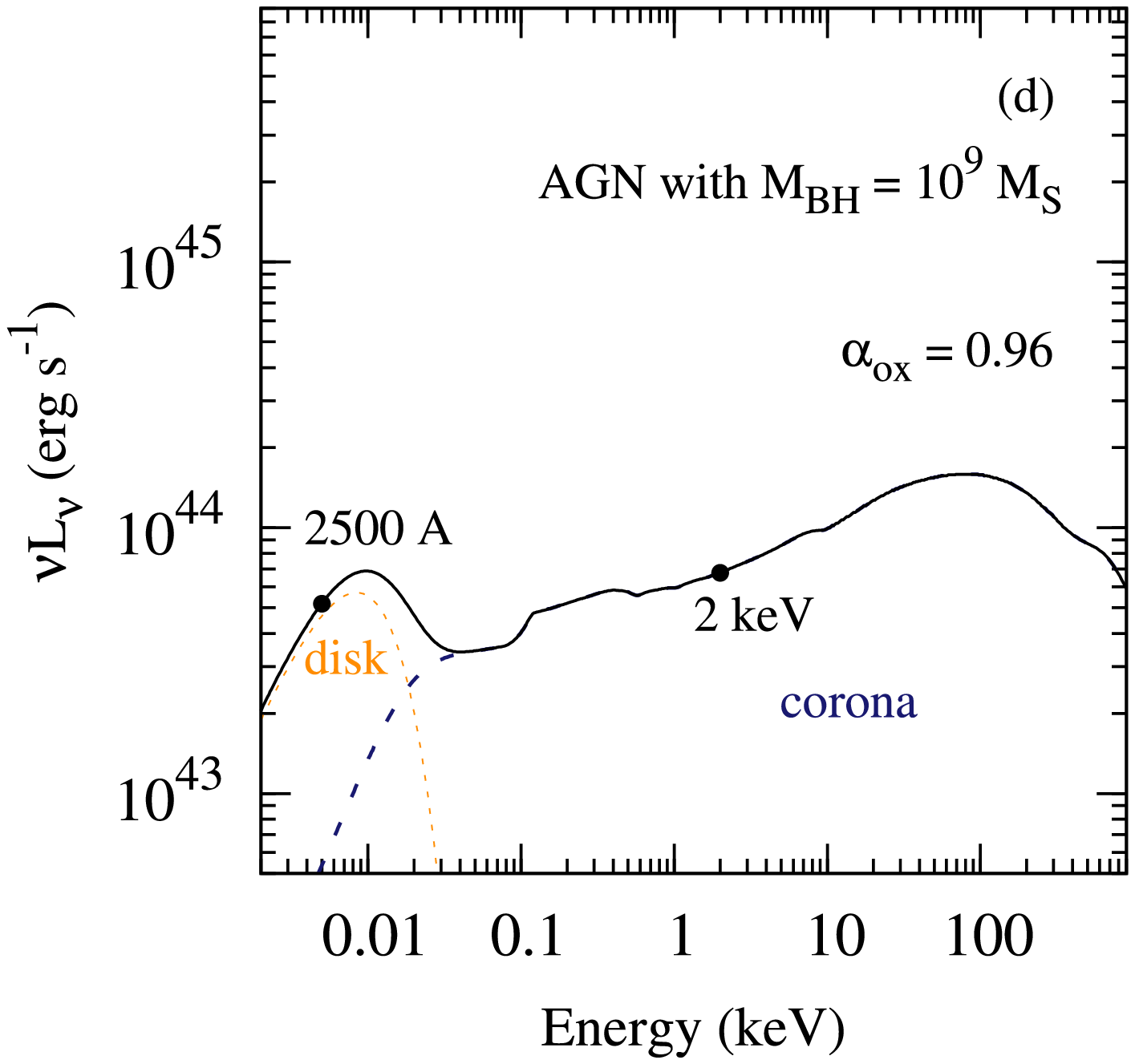}
\caption{(a/c) A soft/hard state spectrum of GRO~J1655-40 (solid/black) formed by a disc blackbody (dotted/orange)
and Comptonization (dashed/blue) components. The 'stretched' disc-to-Comptonization index,
$\alpha^{ \prime}_{\rm GBH}$, was calculated between 3 and 20 keV. (b/d) A soft/hard
state SED of GRO~J1655-40 from (a/c) scaled to the case of an AGN hosting a $10^9$M$_{\odot}$
black hole. The X-ray loudness $\alpha_{\rm ox}$ was calculated between 2500\AA\ and 2 keV.}
\label{fig:sed}
\end{figure*}

\section{GBH template}
\label{sec:data} 

\subsection{Data selection}

\begin{figure*}
\centering
\includegraphics[height=4.3cm,bb=70 60 500 480,clip]{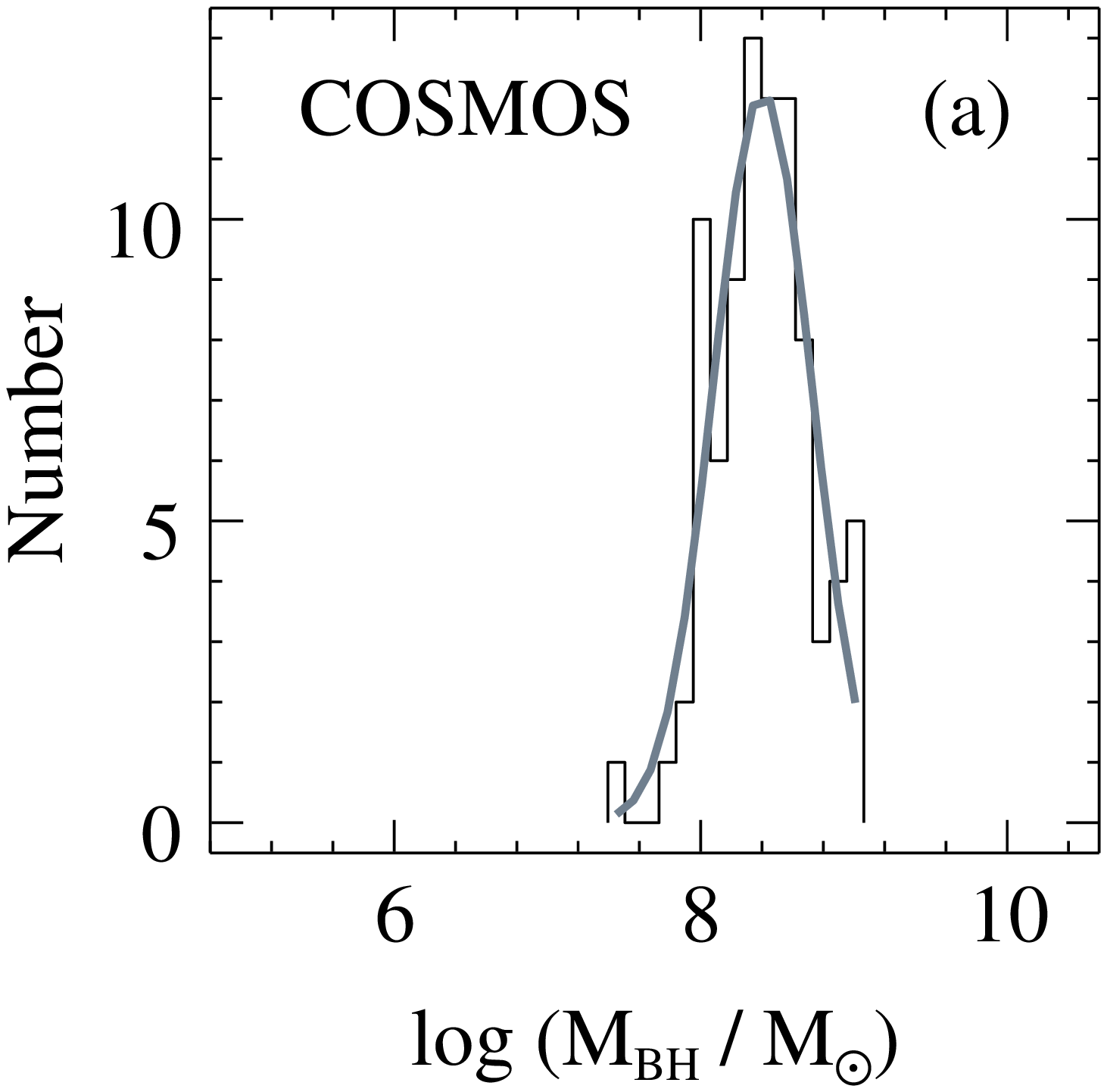}
\includegraphics[height=4.3cm,bb=60 60 500 480,clip]{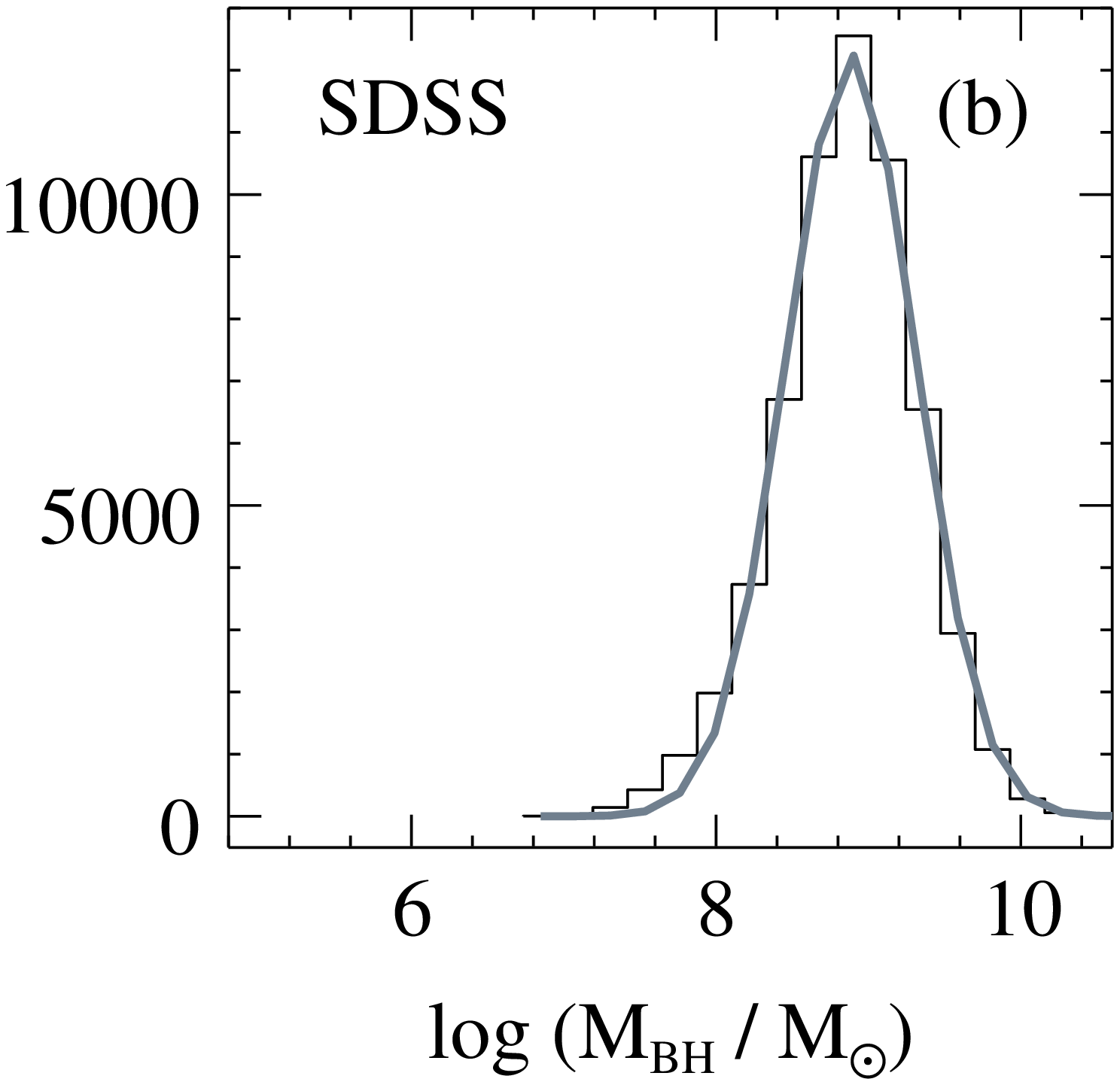}
\includegraphics[height=4.3cm,bb=100 60 500 480,clip]{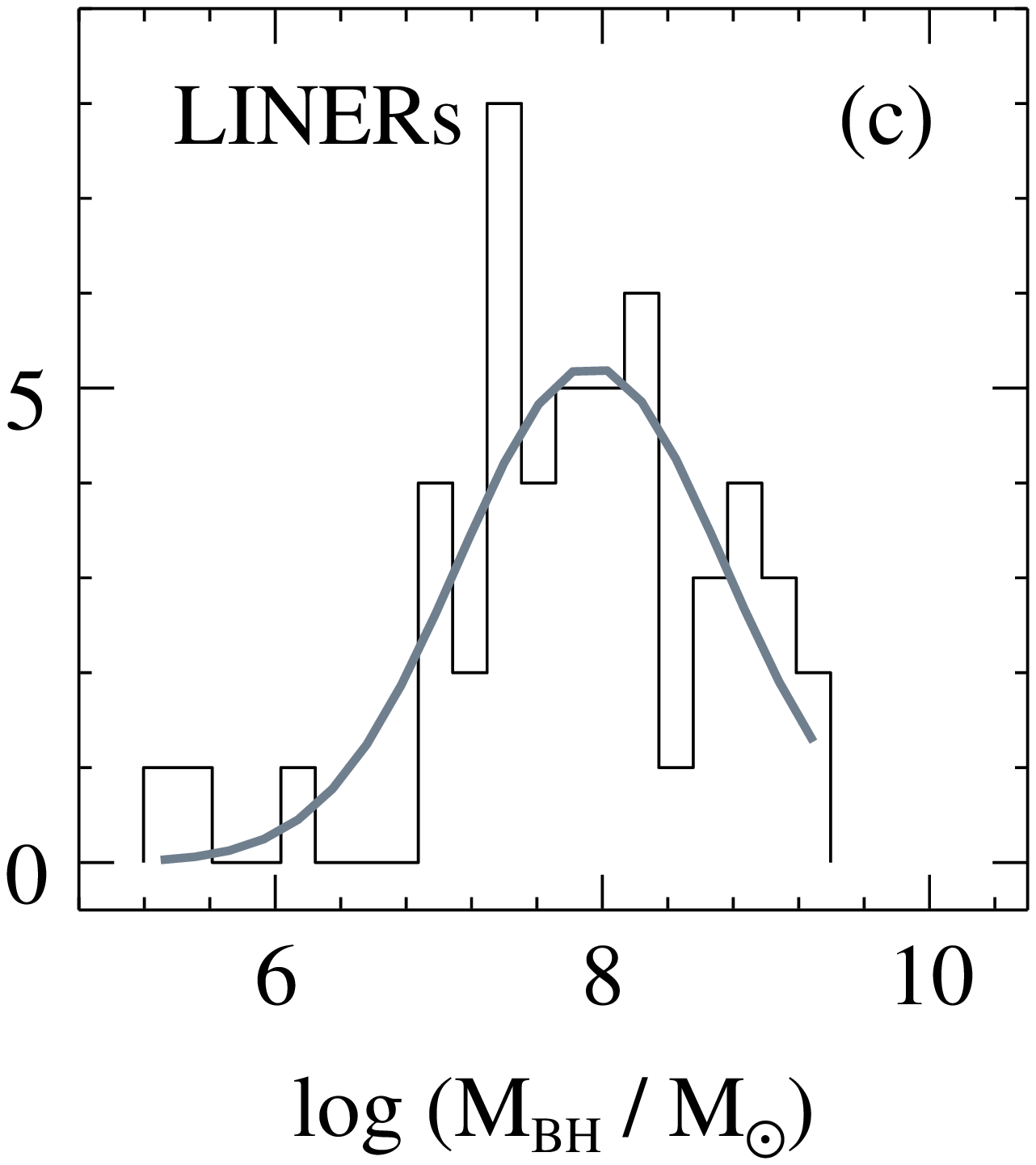}
\includegraphics[height=4.3cm,bb=100 60 500 480,clip]{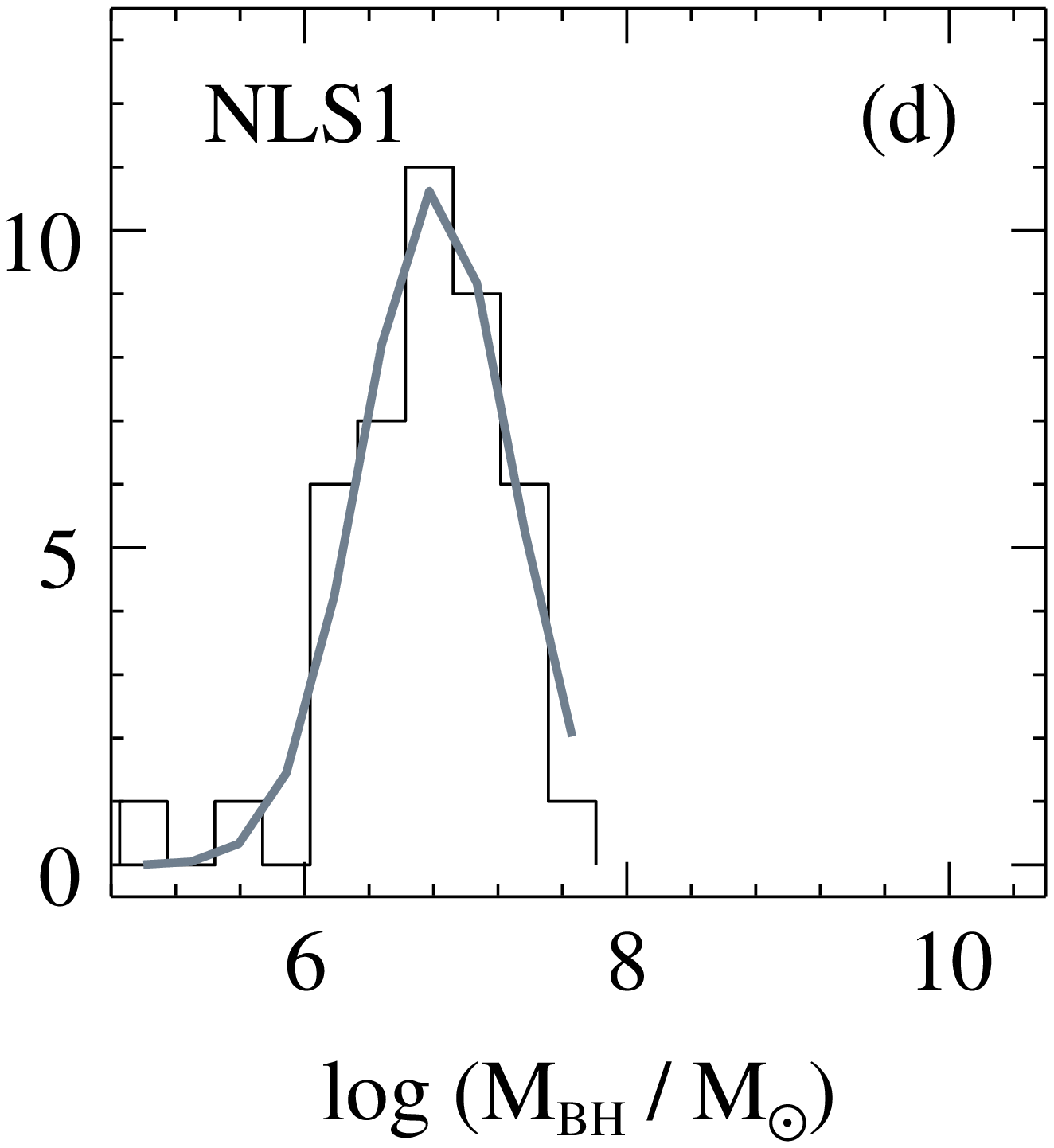}
\caption{Distributions of AGN black hole masses in different observational samples. (a) COSMOS Type 1 AGN from
Merloni et al. (2010), (b) ~60.000 SDSS AGN from Shen et al. (2008), (c) compilation of LINERs from Maoz (2007), Gu \& Cao (2009) \&
Eracleous et al. (2010), (d) NLS1s from Grupe
et al. (2010). The solid curves are log-normal fits to the distributions
with properties listed in Tab.~\ref{tab:tab1}.}
\label{fig:masses}
\end{figure*}

GBH binaries undergo violent events during which their luminosity may increase by more than 3 orders of magnitude.
During these outbursts the sources show variety of different X-ray spectral shapes (e.g. Remillard \& McClintock 2006;
Done et al. 2006). This spectral variability is relatively well understood (e.g. Gierli\'nski \& Done 2004).
Typical peak outburst luminosities do not exceed 0.3--0.5$L/L_E$ (e.g. Done et al. 2006). This is
most probably connected to the physics of the accretion discs that become unstable at high $L/L_E$
(e.g. Merloni 2003, Merloni \& Nayakshin 2006). A few GBHs sporadically reach the high Eddington ratios (see e.g. references
in Done et al. 2006), but only GRS 1915+105 accretes steadily at Eddington ratios $0.3\leq L/L_E \leq 3$. It was
shown in Done, Wardzi\'nski, Gierli\'nski (2004), however, that GRS~1915+105 occupies a different region in their
color-color diagram than the 'normal' GBHs, and it never goes into a 'normal' hard or soft state.

In our study we are looking for a well monitored outburst of a {\it typical} GBH, that would display a
variety of different spectral states, and could be easily described with a standard disc+corona model.
For these reasons, we have selected the 2005 outburst of a confirmed Galactic X-ray black hole
binary GRO~J1655--40 hosting a 6.3M$_{\odot}$ black hole, located at the distance of
3.2 kpc, with inclination of 30$^\circ$ (Remillard \& McClintock 2006). The lightcurve of the
outburst is presented in Fig.~\ref{fig:loop}a.
The X-ray bolometric luminosity (calculated based on the extrapolated 0.01--1000 keV flux of the best fitting models)
changed by more than 3 orders of magnitude, and at the peak of the outburst the system reached
$\sim$20--30\% of the Eddington luminosity.

In 2005 GRO~J1655-40 displayed a variety of GBH spectral states (e.g. Brocksopp et al. 2006; Done et al. 2007),
with the spectral variations covering a typical range for its class. This
can be seen e.g. in the patterns followed by GRO~J1655-40 (Fig.~\ref{fig:loop}b)
and other GBHs (Sobolewska et al. 2009) in the $\alpha_{\rm GBH}$ vs. monochromatic
luminosity at 3 keV diagrams, and also in the hardness-intensity diagrams presented for other GBHs
(e.g. Fender, Belloni \& Gallo 2004, Belloni et al. 2005).
Hence, we argue that the 2005 data of GRO~J1655-40 provide a reasonable SED template to study accretion mechanisms
onto a black hole at various luminosity levels. We note, however, that the hard-to-soft state
transition in 2005 outburst of GRO~J1655-40 happened at $\sim0.02L/L_{\rm E}$, while it was shown
that such a transition may also happen in GBH at much higher ratios, $\geq$0.3$L/L_{E}$ (Gierli\'nski \& Newton 2006).
Thus our template may lack luminous hard state SEDs, and we discuss its implications later on.

We reduced all data of the 2005 outburst found in the public HEASARC\footnote{High Energy
Astrophysics Science Archive Research Center} using {\sc
ftools} ver. 6.2. We extracted PCA\footnote{Proportional Counter Array} spectra for
detector 2, top layer only and HEXTE\footnote{High Energy X-ray Timing Experiment}
spectra from both clusters; one spectrum per pointed observation (see Paper I
for the details of data reduction). For modeling we used 
{\sc xspec} ver. 11.3 (Arnaud 1996).

\subsection{Spectral model}

It has been shown (e.g. Gierli\'nski et al. 1999, Done \& Gierli\'nski 2003) that the spectral
evolution of GBH during an outburst can be explained in
terms of variations of the soft photons temperature and
hard-to-soft compactness ratio, which depends mostly on the geometry of the disc-corona system.
Following these authors, we described the spectra with a sum of the disc blackbody component ({\sc diskbb})
and Comptonization ({\sc eqpair}, Coppi 1999). The {\sc eqpair} code
gives the ratio between the power in seed photons,
$\ell_s$ and hot electrons, $\ell_h$. This ratio,
$\ell_h/\ell_s$, depends mostly on the geometry of the accretion
flow and defines the spectral shape of the hard X-ray continuum. Typically,
the hard state is characterized by $\ell_h/\ell_s \gg 1$, (ultra)
soft state by $\ell_h/\ell_s \ll 1$ and very high/intermediate
state by $\ell_h/\ell_s \sim 1$. The other important model parameters include the 
optical depth, $\tau$, the temperature of the seed photons, $kT_s$, and the ratio of
the nonthermal-to-thermal compactness $\ell_{nth}/\ell_{th}$. The index of the
injected non-thermal electrons was fixed at $\Gamma_{\rm inj} = 2.5$ in all but 14 cases
for which the fits resulted in $1 < \Gamma_{\rm inj} < 3$. The model is not very sensitive
to the changes of the total compactness $\ell_{tot} = \ell_h+\ell_s$.
The {\sc eqpair} also accounts for a reflection of the hard X-rays from a cold
medium (presumably an accretion disc). This reflected component is
parametrized by the reflection amplitude and ionization parameter.
The complete model used in {\sc XSPEC} was defined as {\sc
constant*wabs(diskbb+eqpair)}, where {\sc constant} allows for normalization
between PCA and HEXTE data and {\sc wabs} models the Galactic absorption with $N_H$
fixed at 0.8$\times 10^{22}$cm$^{-2}$ (Done \& Gierli\'nski 2003, and references therein).
The {\sc eqpair} model is particularly well suited for
our study because it allowes for straight forward scaling
of the spectra between the accreting black holes of different mass by applying due
corrections to the seed photons temperature, while keeping the geometry of
accretion constant.

We modeled 94 hard state spectra of the outburst and 38 representative soft state spectra;
132 data sets in total. The times of the data sets
that we used are indicated in Fig.~\ref{fig:loop}a. We set a lower limit of 0.4 keV on the
disc photons temperature, $kT_{bb}$, and in the case of the hard state spectra we fix
$kT_{bb}$ at 0.4 keV. Fits to 119 data sets (90\%)
result in the null hypothesis probability, $p_{\rm null}$, greater than 5\%. The remaining
13 data sets with $p_{\rm null}>5$\% show residuals around 6--7 keV which are
reminiscent of an iron line emission, which is not computed together with the
reflected component by the {\sc eqpair} code. Thus in these 13 cases we include
explicitly the relativistic line model of Laor (1991), {\sc laor}. We fix the
emissivity of the line at 3, and the inner and outer radius at the values used to
compute the reflected component in {\sc eqpair}: 20 and 1000 gravitational radii,
 respectively. After addition of the line, the fits to 9 of 13 data sets resulted in
$p_{\rm null}>5$\%.

We notice that it has been reported by e.g. Sala et al. (2007) and Miller et al.
(2008) that XMM-Newton and Chandra spectra of GRO~J1655-40, respectively, show
features implying strong accretion disc winds. A number of absorption lines in the
7--9 keV band have been observed. However, RXTE has a much poorer energy resolution
than Chandra or XMM-Newton, and so we are not able to handle properly these
absorption features.

Nevertheless, we conclude that we have found a model that fits well 128 of 132
considered data sets, which means that the model fits our set of data well from the
statistical point of view (see e.g. Sobolewska \& Papadakis 2009). An example
of soft/hard state models fitted to the GRO~J1655-40 data is shown in
Figs.~\ref{fig:sed}a/c.

Following Paper I, we calculate the disc-to-Comptonization index, $\alpha^{\prime}_{\rm GBH}$, for
GRO~J1655-40 based on the best fitting spectral models. The diagram
of $\alpha^{\prime}_{\rm GBH}$ versus the $(ef_e)_3$ flux in keV cm$^{-2}$ s$^{-1}$ at 3
keV is presented in Fig.~\ref{fig:loop}b, and the hard/soft spectral states that GRO~J1655-40
entered during its evolution are indicated with different colors. The values
of $\alpha^{\prime}_{\rm GBH}$ for GRO~J1655-40 range between 1 and 2, which is in
general true also in the case of other GBH (see Paper I), and in the case of AGN's $\alpha_{\rm ox}$.

\section{Method}
\label{sec:method}

\begin{figure*}
\centering
\includegraphics[height=5.8cm,bb=165 180 535 580,clip,angle=-90]{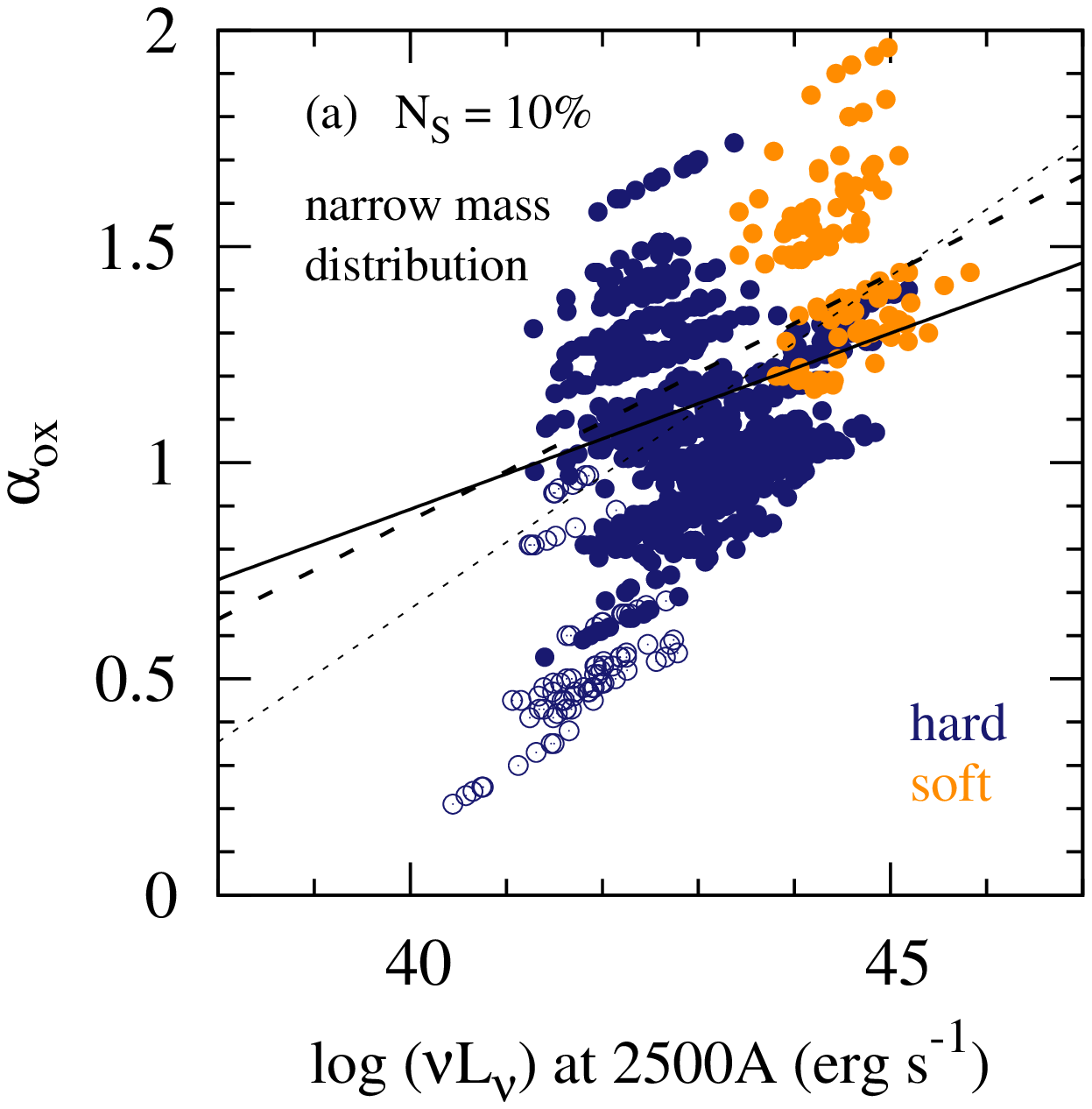}
\includegraphics[height=5.8cm,bb=165 180 535 580,clip,angle=-90]{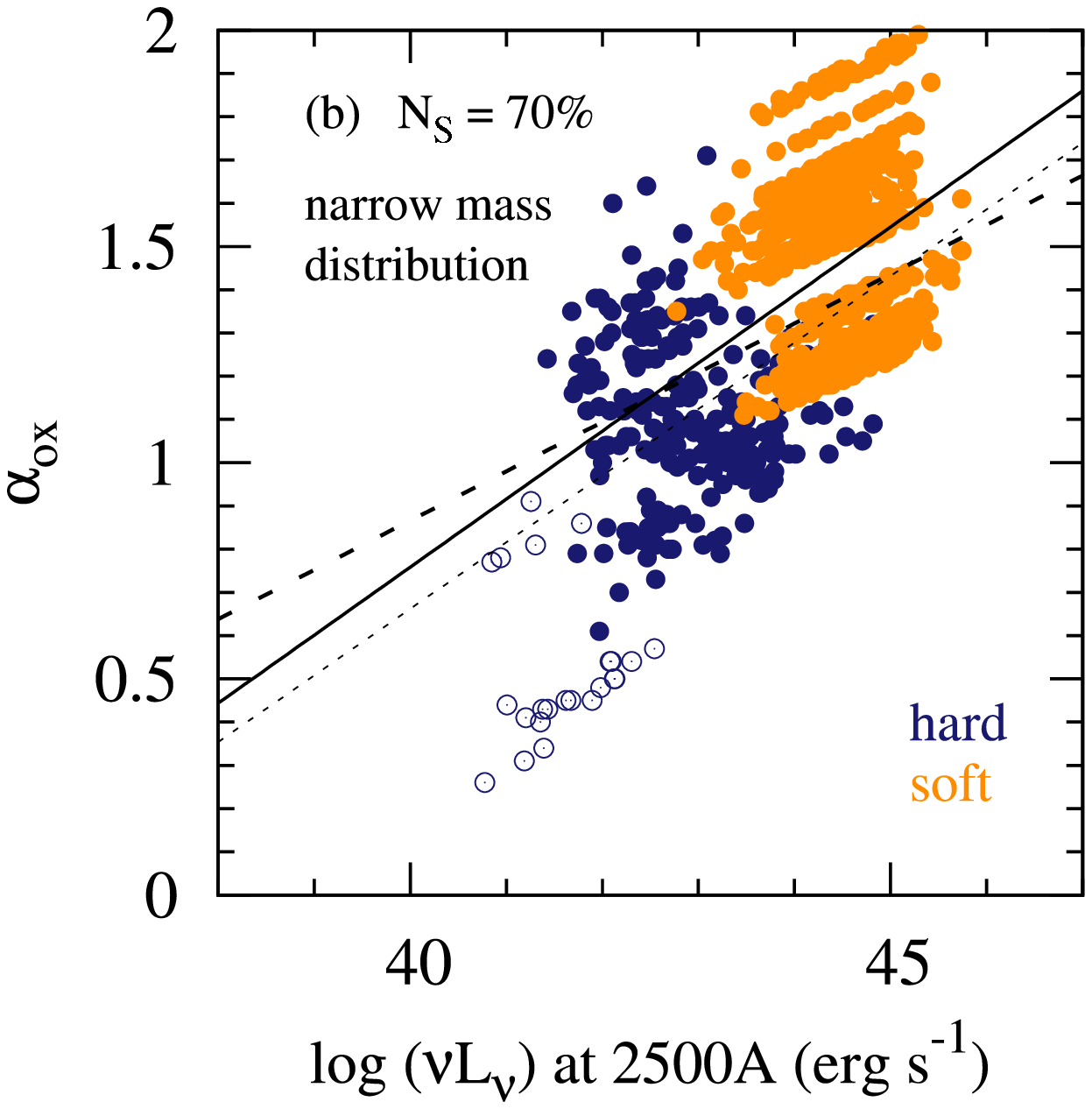}
\includegraphics[height=5.8cm,bb=165 180 535 580,clip,angle=-90]{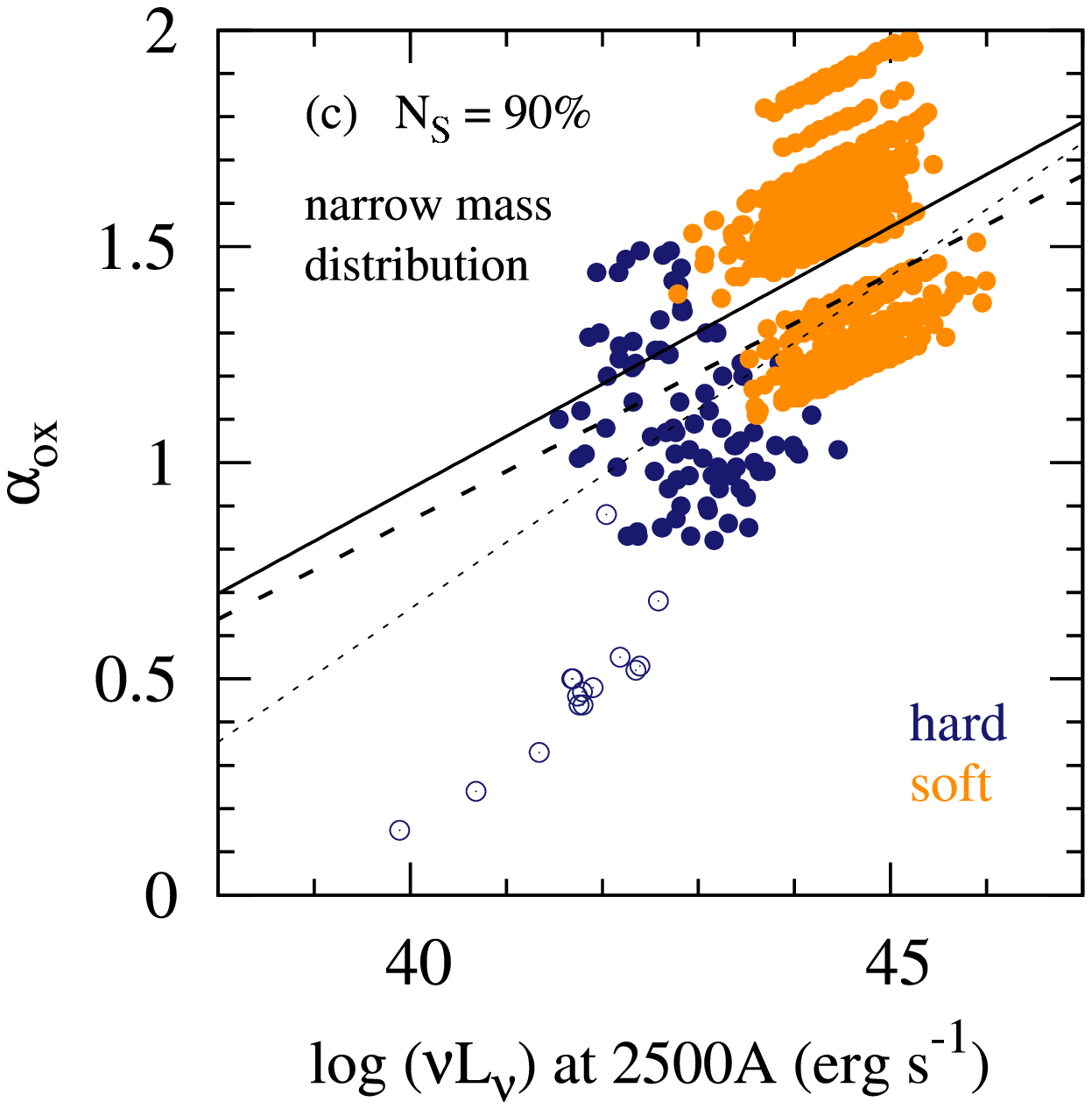}\\
~~~~~~~~~~~~~~~~~~~~~~~~~~~~~~~~~~~~~~~~~~~~~~~~~~~~~~~~~~~~~\includegraphics[height=5.8cm,bb=165 180 535 580,clip,angle=-90]{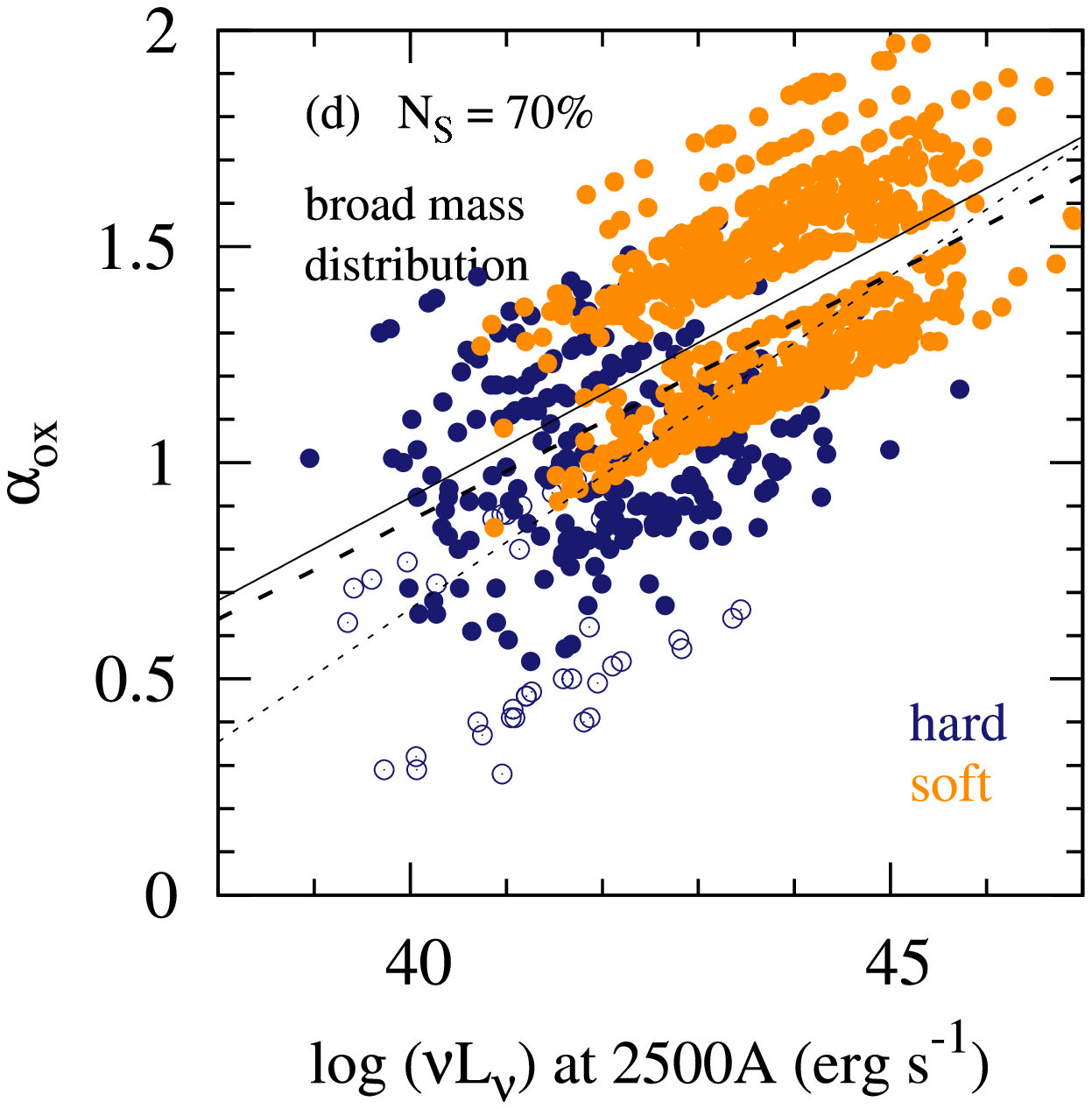}
\includegraphics[height=5.8cm,bb=165 180 535 580,clip,angle=-90]{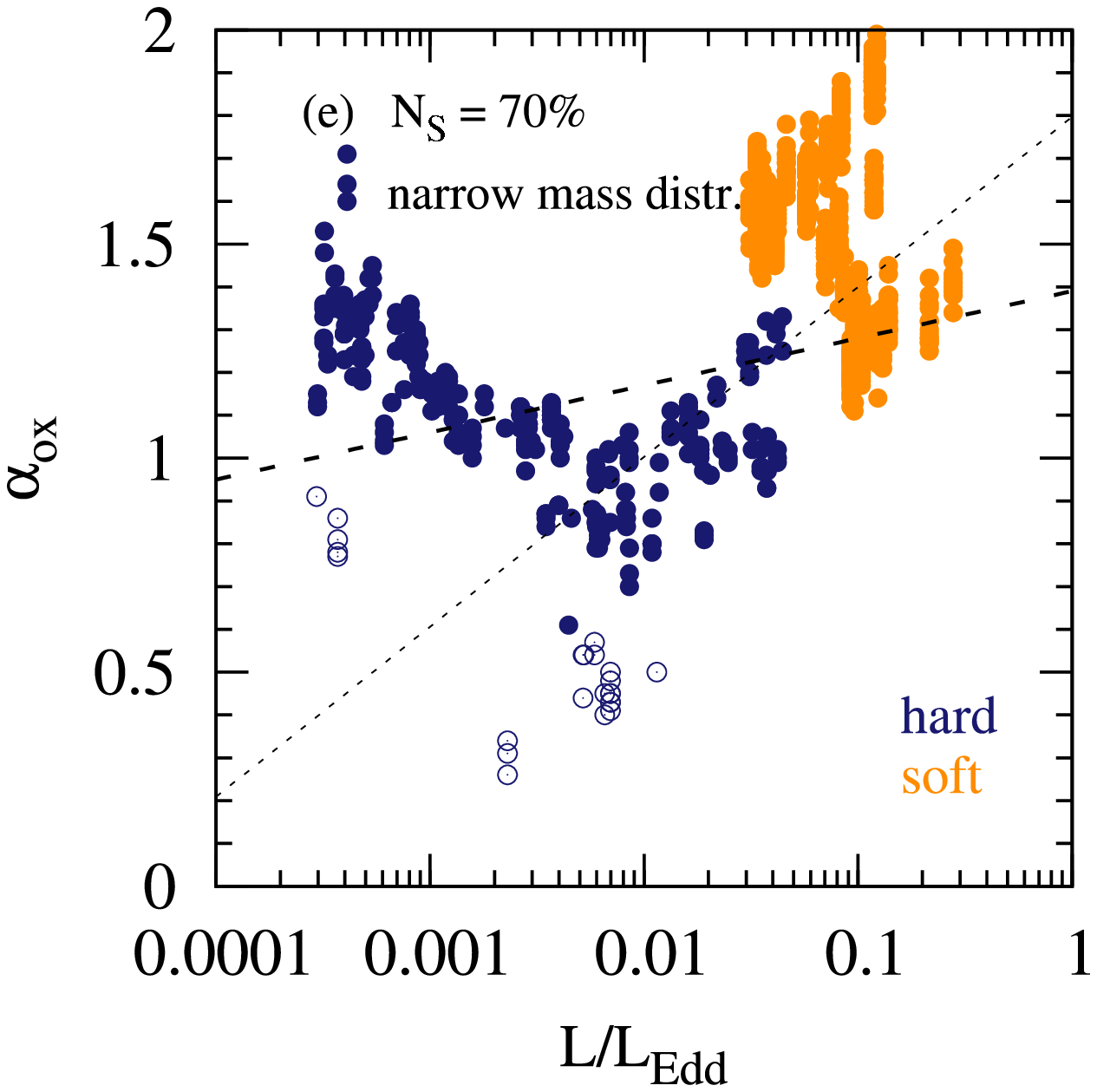}
\caption{Simulated X-ray loudness as a function of simulated monochromatic luminosity
at 2500\AA\/ (a--d) and Eddington luminosity ratio (e). The black hole mass was drawn from the distribution in Fig.
3a (a--c,e) and 3c (d). The number of the soft state AGN was assumed to be $N_S = 70$\% in (b,d--e), 10\% (a) or 90\% (c),
hard states - black/blue, soft state - gray/orange, hard state with
insignificant disc component - open circles. From comparison of panels (b) and (d) it can be seen that for a narrow
FWHM of the AGN mass distribution, the simulated AGN spectral are well separated, and they overlap when the FWHM increases.
The solid lines represent the best fits to the simulations, while the dotted and dashed lines are the best fits to the
observational samples of Lusso et al. (2010; 545 X-ray selected type 1 AGN, from the XMM-COSMOS) and Grupe et al.
(2010; 92 bright soft X-ray selected AGN observed with Swift), respectively. The hard state simulations indicated
with the open circles where neglected during the fit because very low level of the disc component would make such
AGN spectra difficult to recognize in real data. The slope of the simulated correlation depends on the assumed number
of the soft state AGN in the sample. (e) A change of the sign of the $\alpha_{ox}$ vs. $L/L_E$ correlation can be observed for
$L/L_{\rm E} \approx 0.01$. This result does not depend on the adopted AGN mass distribution.
The dotted and dashed lines correspond to the correlations found by Lusso et al. (2010)
and Grupe et al. (2010), respectively.}
\label{fig:sim}
\end{figure*}

We simulate spectra of AGN by scaling the disc temperature and the bolometric
luminosity of our collection of the best-fitting GRO~J1655--40 models. For a standard Shakura-Sunyaev
geometrically thin optically thick accretion disc, the disc temperature scales
with the black hole mass as $T_{\rm disc} \propto M^{-1/4}$, and
the bolometric luminosity is proportional to the black hole mass, $L_{\rm bol} \propto
M$. We assume that the accretion flow geometry, and hence the heating to cooling
compactness ratio, $\ell_h/\ell_s$, varies in the same way as the function of the mass
accretion rate both in Galactic and supermassive black holes. We set the normalization of the iron line,
when present in the baseline spectrum, to zero, since it does not contribute significantly to the total simulated AGN flux.
Our goal is to check if taking into account the mass distribution among the AGN samples can provide insights
to the origins of the observed dependence between the X-ray loudness, $\alpha_{\rm ox}$)
and optical/UV luminosity at 2500\AA.

\subsection{Masses of AGN}

For our study we select the AGN samples with measured black hole mass. Black hole
mass measurements in AGN are not trivial. However, various methods (e.g. reverberation mapping and
correlations with the widths of emission lines) resulted in
black hole mass estimates ranging between 10$^6$ and 10$^{10}$ Solar
masses. Figure \ref{fig:masses}
shows mass distributions for samples of AGN compiled from the literature.

\begin{table}
\caption{Properties of lognormal mass distributions in different AGN samples from
literature.}
\begin{center}
\begin{tabular}{l c c c l}
\hline
Sample & \#of AGN & $<\log \frac{{\rm M}_{\rm BH}}{{\rm M}_{\odot}}>$ & FWHM & Ref.\\
\hline
T1 COSMOS & 86           & 8.40 & 0.75 & [1]   \\
T1 SDSS    & $\sim$60.000 & 8.89 & 1.00 & [2]  \\
LINERs     & 50           & 7.9  & 1.9  & [3--5] \\
NLS1       & 42           & 6.8  & 1.1  & [6]  \\
\hline
\end{tabular}
\end{center}
[1] Merloni et al. (2010), [2] Shen et al. (2008), [3] Maoz (2007), [4] Gu \& Cao
(2009), [5] Eracleous, Hwang \& Flohic (2010), [6] Grupe et al. (2010).
\label{tab:tab1}
\end{table}

In Fig.~\ref{fig:masses}a we show
the mass distribution reported for 86 Type 1 Broad Line AGN by Merloni et al. (2010) as part of the COSMOS project.
The masses were consistently calculated based on the MgII emission line, however the sample spans
relatively narrow ranges of redshifts ($1<z<2.2$) and luminosities ($44.5<\log L_{\rm bol} < 46.5$).
For comparison, in Fig.~\ref{fig:masses}b we show the results of Shen et al. (2008), for Type 1 Broad Line AGN in SDSS.
This sample contains $\sim$60.000 objects with redshifts and luminosities in the $0.1<z<4.5$ and $44.5<\log L_{\rm bol} < 48$ range,
with masses calculated based on MgII, CIV or H$_{\beta}$ lines. This is the largest AGN sample with black hole mass
measurements. However, the authors point out that the estimations based on the CIV line may be biased due to the
line being disc wind dominated.

In addition to the Type 1 Broad Line AGN, we consider a sample of LINER galaxies. Black hole mass
measurements are challenging because of the host galaxy domination in the optical band. We compile 
a LINERs sample galaxies from Maoz (2007), Gu \& Cao (2009) and Eracleous et al. (2010)
containing black hole mass estimates for 50 sources. The estimated Eddington ratios for LINERs
(e.g. Gu \& Cao 2009, Eracleous et al. 2010) indicate that they accrete at low rates
and thus may be the counterparts of the hard state GBHs. We test this possibility in Sec. 4.2.

Finally, we consider Narrow Line Seyfert 1 galaxies (NLS1). We study a NLS1 subsample from Grupe et al. (2010)
defined as FWHM(H$_{\beta}) < 2000$ km s$^{-1}$ (Osterbrock \& Pogge, 1985; Goodrich 1989). These objects have
reliable estimates of the accretion rates based on the simultaneous SEDs in optical and X-ray bands. NLS1 are
believed to have higher Eddington ratios (and lighter black holes) than their broad line counterparts.
In Sec. 4.2 we study if they can be considered as counterparts of the soft state GBHs.

Distribution of AGN masses in all the above samples can be satisfactorily described
by a lognormal functions with means and FWHMs listed in Table \ref{tab:tab1}. Thus,
in this paper we will assume that the masses of black holes in AGN have a lognormal
distribution. The means of the above distributions are different: $\log ({\rm M}_{\rm BH}/{\rm M}_{\odot}) = 6.8$
and 8.9 for NLS1s and SDSS samples, respectively. The FWHMs vary between 0.75 and 1.9 for the COSMOS and LINERs samples,
respectively. We will study the effect of a narrow vs. broad mass distribution in our simulations.

\subsection{Spectral states of AGN}

Based on the modelling of the GRO~J1655-40 spectra we conclude that this source was in a soft spectral state
for $t_s\sim0.7 {\rm T}$, and in a hard spectral state for $t_h\sim0.3 {\rm T}$ of the 2005 outburst (where
T is the total time of the outburst). The situation in any real unbiased AGN
sample would be different in the sense that in the case of GRO~J1655-40 we have many spectra of the same source in
various spectral states, while in a real AGN sample we would record usually one spectrum for each AGN. However,
a large AGN sample would contain a mixture of hard and soft state SEDs, with a fraction of the soft state SEDs,
$N_S/N$ (where $N_S$ and $N$ are the number of the soft AGN SEDs and the total number of AGN SEDs, respectively),
corresponding to $t_s/{\rm T}$ in a typical GBH outburst.

The duration of the time spend in either state differs for different GBHs (see e.g. Figs. 4--5 in Done et al. 2007),
and depends most probably on a number of physical parameters. It is not clear
what would be the expected ratio of the hard to soft state AGN (or GBH) spectra in a large sample. Thus, we 
first assume that $N_S/N = 0.7$ (corresponding to the $t_s/{\rm T}$ ratio in the 2005 outburst
of GRO~J1655-40), and then we check how our results change if we assume that $N_S/N = 0.1$ and 0.9.

To assign a spectral state to the simulated AGN spectrum we draw a number from a uniform
distribution between 0 and 1. If this number is between 0 and $N_S/N$, we
assign a soft spectral state to the AGN spectrum that will be simulated. Otherwise the AGN
spectrum is simulated assuming that the spectral state is hard.

\begin{figure*}
\centering
\includegraphics[height=5.8cm,bb=165 180 535 580,clip,angle=-90]{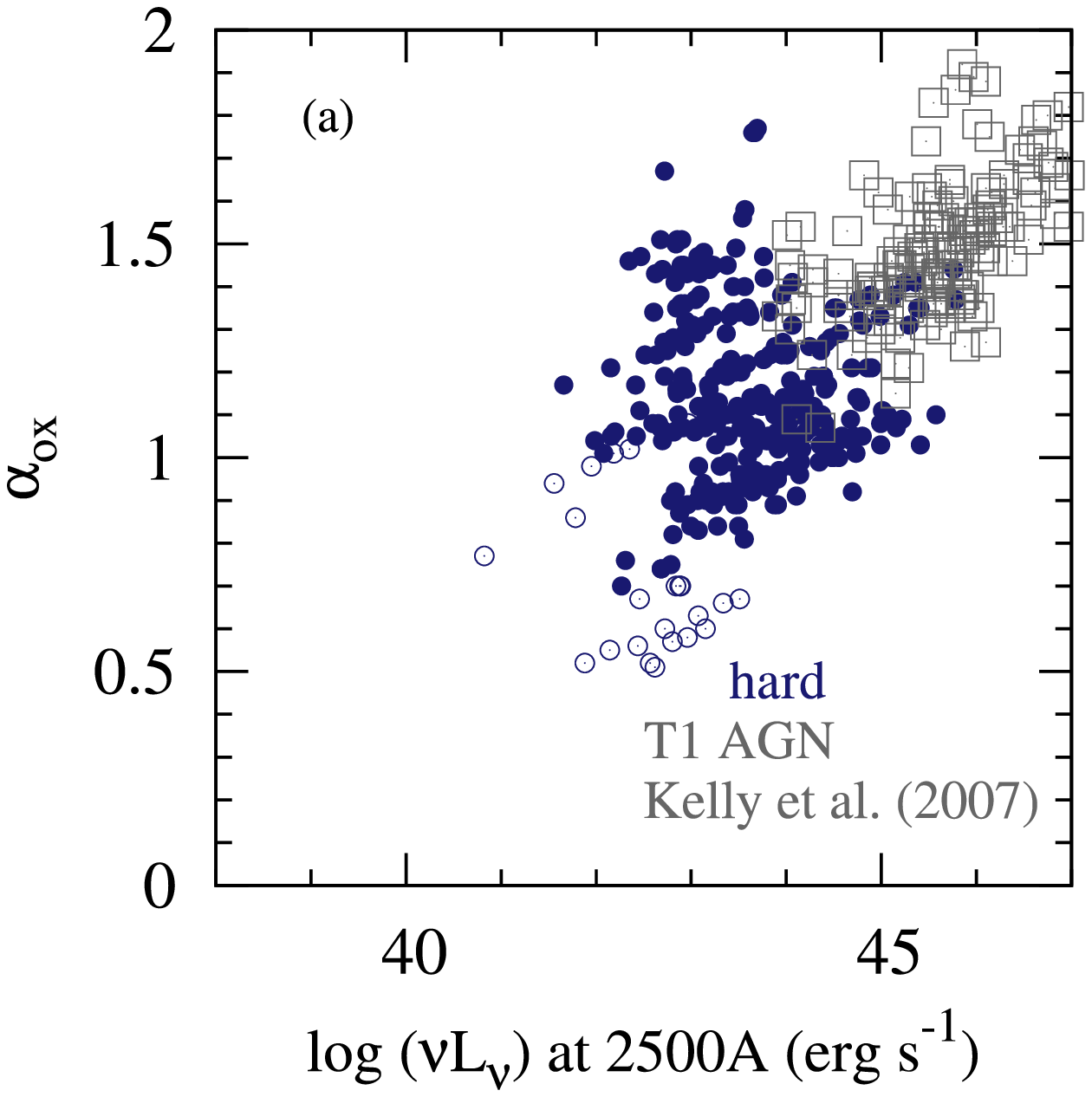}
\includegraphics[height=5.8cm,bb=165 180 535 580,clip,angle=-90]{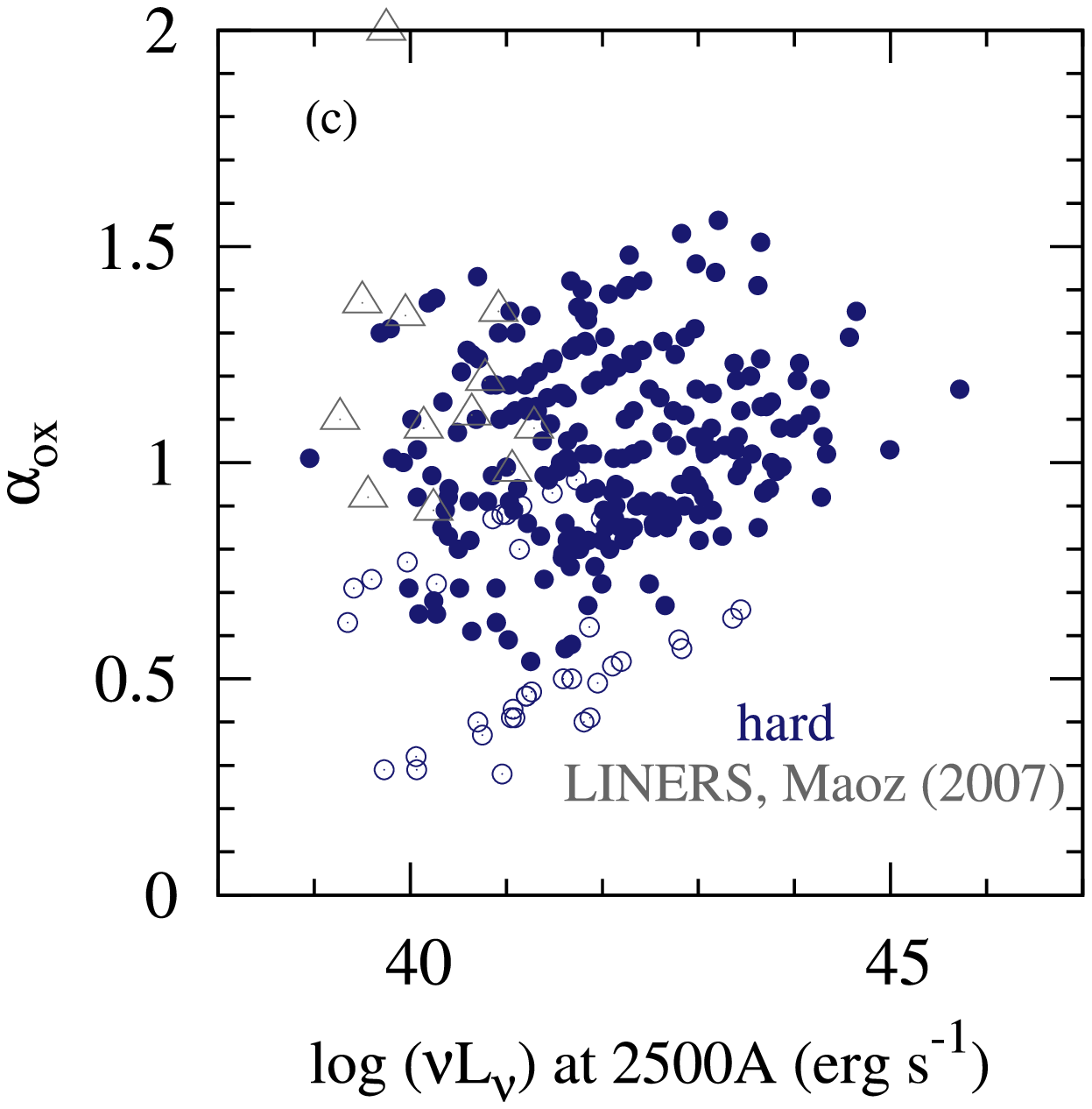}
\includegraphics[height=5.8cm,bb=165 180 535 580,clip,angle=-90]{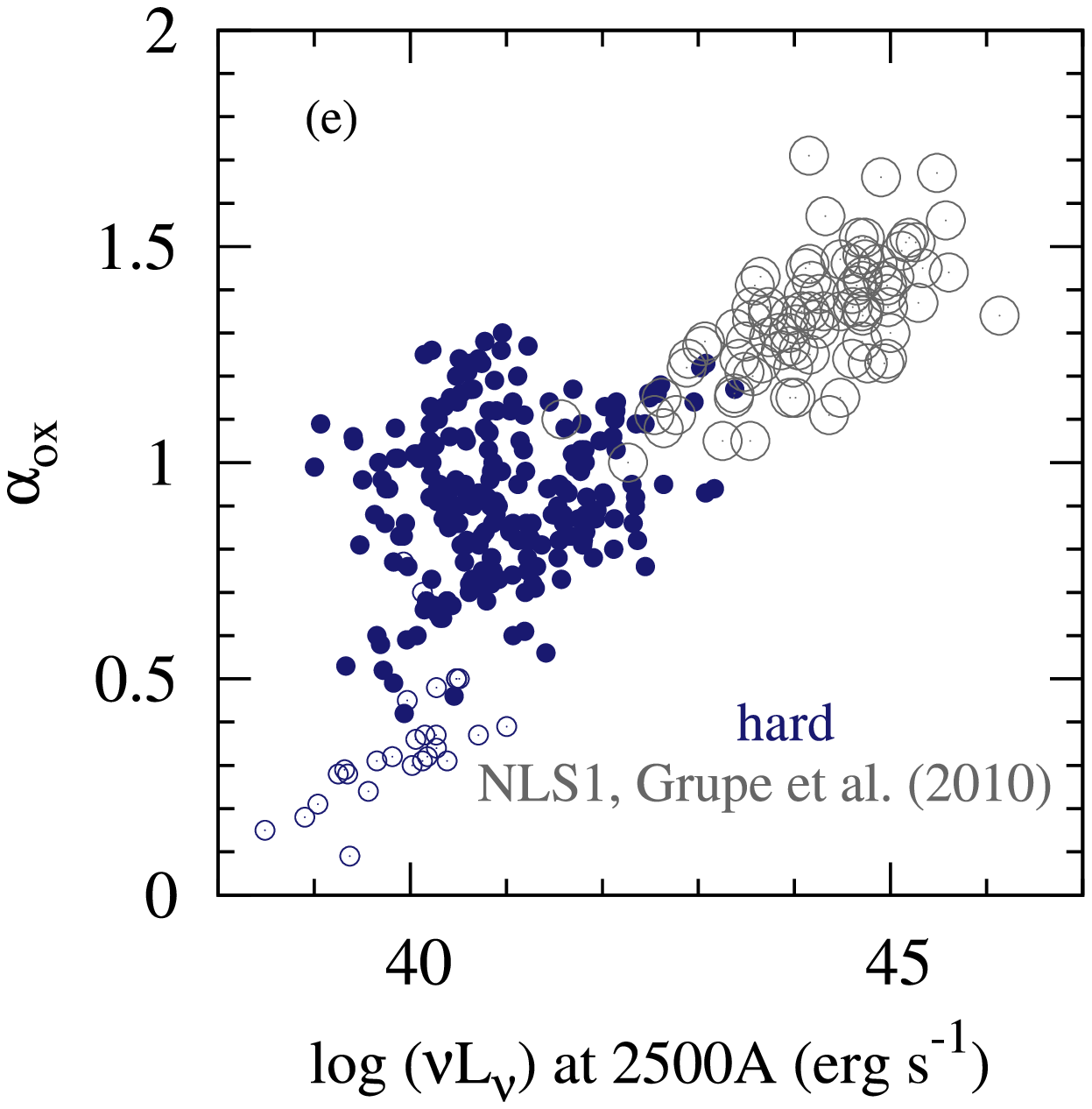}\\
\includegraphics[height=5.8cm,bb=165 180 535 580,clip,angle=-90]{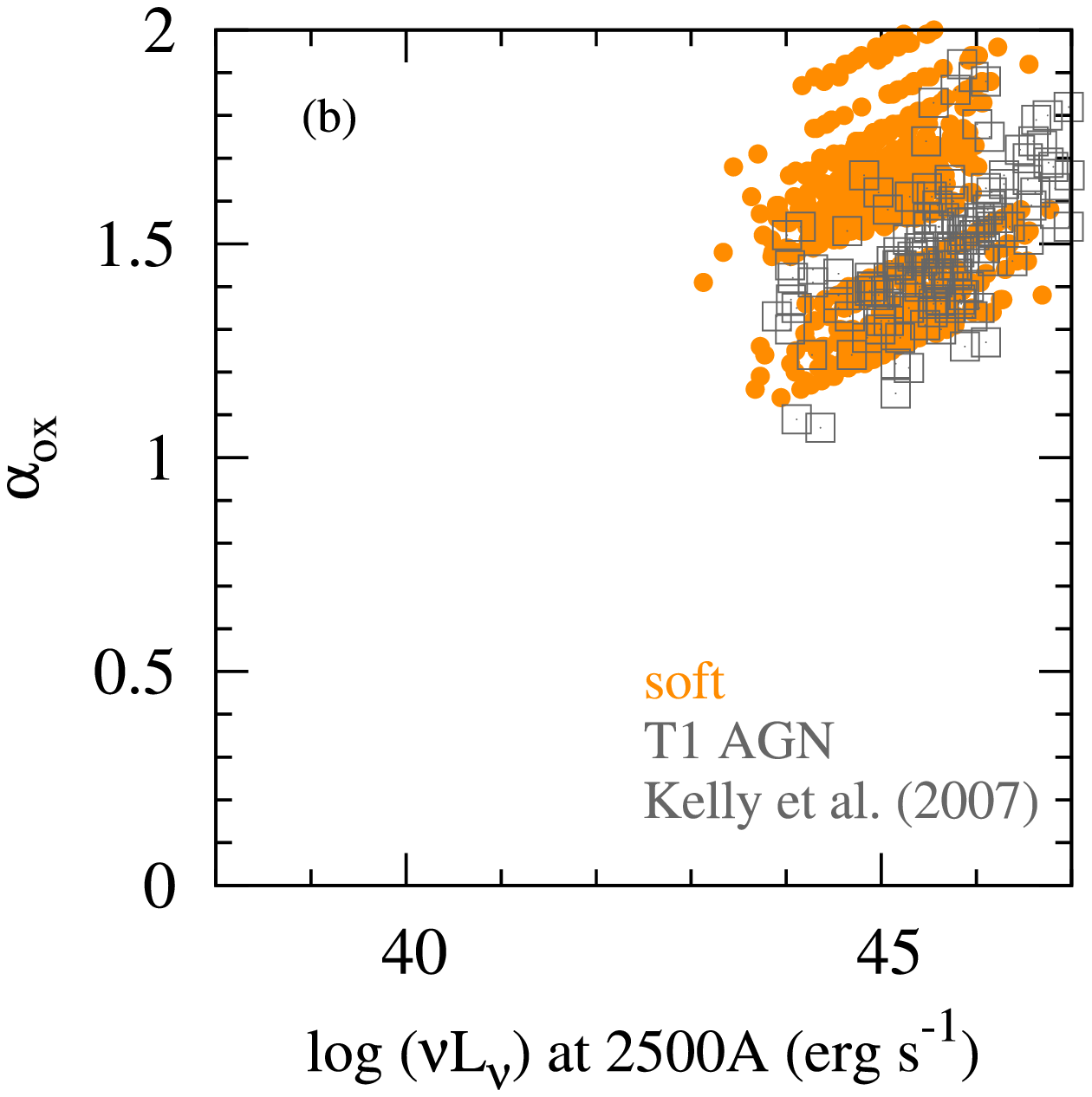}
\includegraphics[height=5.8cm,bb=165 180 535 580,clip,angle=-90]{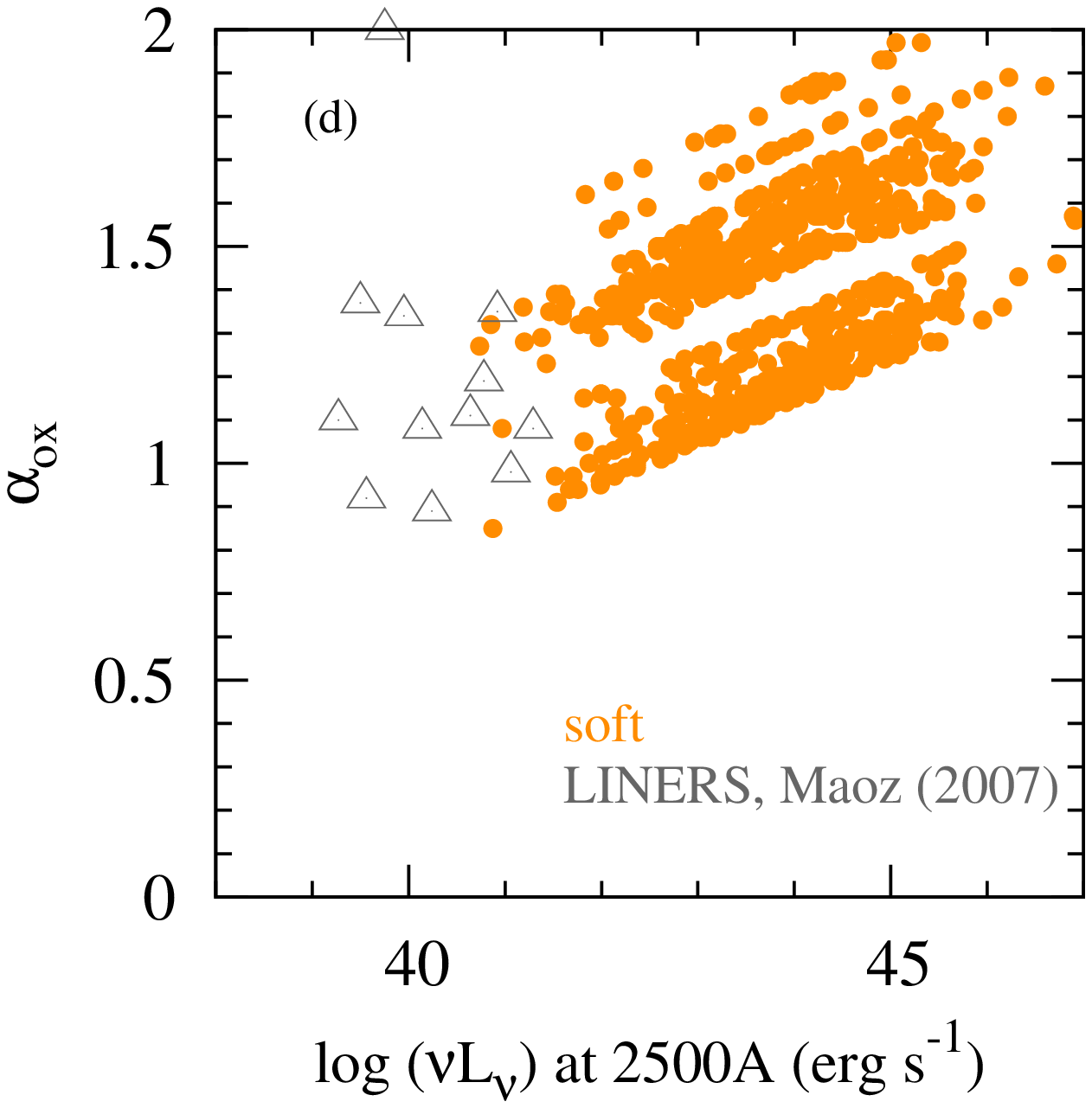}
\includegraphics[height=5.8cm,bb=165 180 535 580,clip,angle=-90]{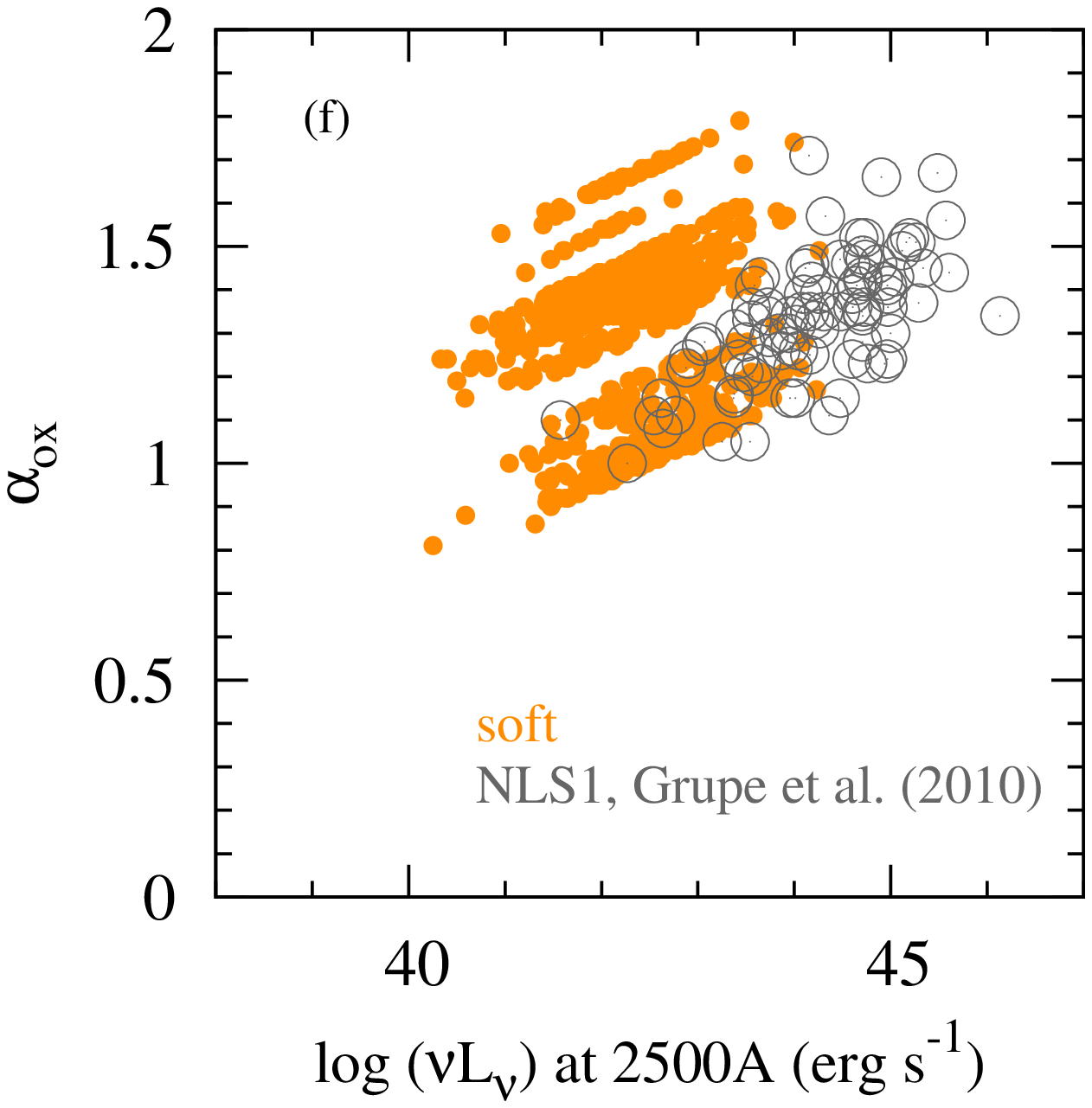}
\caption{Comparison of properties of the simulated AGN spectra with observational
samples, hard states - black/blue, soft state - gray/orange, hard state with
insignificant disc component - open circles. (a-b) Kelly et al. (2008) Type 1 AGN,
(c-d) Maoz (2007) LINERs, (e-f) NLS1 from Grupe et al. (2010) superimposed on the simulated hard and soft
state AGN data (with $\log$ of AGN mass drawn from a Gaussian
distribution as in Figs.~\ref{fig:masses}a, \ref{fig:masses}c and \ref{fig:masses}d, respectively).
Observed LINERs overlap with the hard state simulated AGN, while observed Type 1 quasars and NLS1s overlap
with the simulated soft state AGN.}
\label{fig:data}
\end{figure*}

\subsection{Simulating the AGN SEDs}

After determining the simulated AGN spectral state, we randomly
choose a spectrum from the 94 hard or 38 soft state observations of GRO~J1655-40
to access the mass accretion rate and
spectral parameters that we assume to be constant between the original GBH and
simulated AGN spectra, and independent on the mass of the accreting black hole:
the $\ell_h/\ell_s$ compactness ratio, the optical depth of the Comptonizing corona,
the seed photons Compactness, and the thermal/non-thermal electron fraction in the
corona. Next, we randomly draw a black hole mass from an observationally motivated lognormal
distribution, and scale the disc temperature and luminosity in the original GBH
spectrum with the mass. We repeat the entire procedure $N=1000$ times to get a large enough
simulated sample of AGN spectra for comparison with the observational data. As an example,
in Fig.~\ref{fig:sed}b/d we plot a simulated soft/hard state AGN spectrum which was obtained
by scaling a soft/hard state GBH spectrum (Fig. \ref{fig:sed}a/c) assuming that the AGN hosts
a 10$^9$M$_\odot$ black hole.

\section{Results}
\label{sec:res}

We calculate $\alpha_{\rm ox}$ for each simulated AGN spectrum and we plot it as a
function of the simulated monochromatic luminosity at 2500\AA. The observed 
correlation is clearly present in the simulated data.
Initially, we investigate the effect of the width of the black hole mass distribution on
the simulated $\alpha_{ox}$ vs. $\nu L_{\nu}$ at 2500\AA\/ correlation. 

First we perform our simulations for the log-normal AGN mass distribution as in Merloni et al. (2010, Fig.~\ref{fig:masses}a),
with the narrowest FWHM=0.75 among the four distributions in Fig.~\ref{fig:masses}. We assume that $N_S = 70$\%
of the simulated AGN are in a soft spectral state.
The results are presented in Fig.~\ref{fig:sim}b. Indeed, observationally
motivated log-normal mass distribution in this sample leads to a correlation between
the $\alpha_{\rm ox}$ and monochromatic optical/UV luminosity at 2500\AA\/. The hard and soft simulated states
are well separated.

Next, we repeat our simulations using the lognormal mass distribution
resulting from fitting the combined Maoz (2007), Gu \& Cao (2009) and Eracleous et al. (2010) LINERs masses (Fig.~\ref{fig:masses}c).
This distribution has FWHM=1.9, the broadest among the four samples in Fig.~\ref{fig:masses}. It can be seen in Fig.~\ref{fig:sim}d
that for LINERs the hard and soft states overlap for $\sim$5 orders of magnitude in $(\nu L_\nu)_o$.
Based on this we conclude that the larger the width of the distribution used to draw the AGN black hole masses,
the more significant the spread of the simulated quantities, and the more pronounced the overlap
of the hard and soft spectral states.

\subsection{Correlation analysis}

{We quantify the simulated $\alpha_{\rm ox}$ vs. $(\nu L_\nu)_o$ correlations by fitting the simulated
data with a function of the following form: $\alpha_{\rm ox} = a \times \log(\nu L_\nu)_o + b$.}

During the fit we neglect the hard state data indicated with open circles which where simulated based on the
GRO~J1655-40 observations with the disc flux lower than $F_{d} = 5\times10^{-11}$ erg s$^{-1}$ cm$^{-2}$.
Among the 94 considered GRO~J1655-40 hard state observations there were 11 datasets that were well
described by the Comptonized continuum with a flux $F_c$ dominating significantly over the disc flux $F_d$ in the total spectrum.
A case of an AGN with similar geometry (i.e. extremely low disc flux) would be difficult to detect in optical/UV
due to the emission of a host galaxy that would dominate in this band. However, such an AGN would still be detected
in the X-ray band with an X-ray SED resembling a power law.

Our fits result in
\begin{eqnarray}
\alpha_{ox}^{\rm COSM,70} = (0.157\pm0.008)\times \log(\nu L_\nu)_o -
(5.5\pm0.4)\nonumber\\
\alpha_{ox}^{\rm SDSS,70}   = (0.161\pm0.008)\times \log(\nu L_\nu)_o -
(5.8\pm0.4)\nonumber\\
\alpha_{ox}^{\rm LINER,70} = (0.119\pm0.005)\times \log(\nu L_\nu)_o -
(3.8\pm0.2)\\
\alpha_{ox}^{\rm NLS1,70}   = (0.125\pm0.008)\times \log(\nu L_\nu)_o -
(4.1\pm0.3)\nonumber
\end{eqnarray}
for AGN SED sets simulated based on all four AGN mass distributions from Fig.~\ref{fig:masses}, respectively.
The range of slopes we find in
this correlation is consistent with that seen in real data. For example, Grupe et al. (2010) found a slope of 
0.114$\pm$0.014 based on the study of 92 bright soft X-ray selected AGN observed with Swift (a combination of
narrow line and broad line Seyfert 1s), while Lusso et al. (2010)
reported a slope of 0.154$\pm$0.010 in the XMM-COSMOS sample of 545 X-ray selected type 1 AGN.  Their correlations
are indicated in Fig.~\ref{fig:sim} with dotted and dashed lines, respectively. The best fits
to the SEDs simulated using narrow (COSMOS) and broad (LINERs) mass distributions are indicated with solid lines in
Figs.~\ref{fig:sim}a--d.

%
%
%

The fitted slopes depend on the assumed number of the soft state spectra in the AGN sample.
Figures~\ref{fig:sim}a and \ref{fig:sim}c show how the correlation changes if we assume
that $N_S = 10$\% or 90\% of the AGN, respectively, are in a soft state, for the case of
a narrow mass distribution from Fig.~\ref{fig:masses}a. For the four considered
AGN mass distributions we obtained $a^{10} = 0.078$--0.85 for $N_S = 10$\% and $a^{90} = 0.107$--0.121 for $N_S = 90$\%.

Additionally, we check if the simulated AGN's $\alpha_{\rm ox}$ is related to the bolometric
luminosity in Eddington units. In Fig.~\ref{fig:sim}e we plot $\alpha_{ox}$ vs. $L/L_{E}$ for a narrow mass
distribution (Fig.~\ref{fig:masses}a), assuming $N_S = 70$\%.
Interestingly, the simulated X-ray loudness correlates positively with the Eddington ratio down to approximately
$\lambda_{\rm crit} \approx 0.01$ $L/L_{\rm E}$ and below this value the correlation changes its sign.
In Fig.~\ref{fig:sim}c we superimpose our simulations and the correlations found in real data by Lusso et al. (2010)
and Grupe et al. (2010) (dotted and dashed lines, respectively). The correlation reported by Lusso et al. matches the slope
produced by our simulated SEDs down to the critical Eddington ratio, $\lambda_{\rm crit}$. These results
do not depend either on the properties of the adopted mass distribution or on $N_S$.

\subsection{Spectral states of AGN}

We compare our simulated data with the known AGN samples, which
have measurements of both the monochromatic luminosity at 2500\AA\/ and $\alpha_{\rm ox}$, and
preferably also the black hole estimates.
We assume in this section that $N_S = 70$\%, but we stress that the results do
not depend on this parameter.

Neither Merloni et al. (2010) nor Shen et al. (2008) have X-ray observations for their complete samples.
For that reason, first we consider the broad line AGN (a SDSS subsample) from Kelly et al. (2007).
We compare them with the simulated data resulting from
hard (black/blue) and soft (gray/orange) state spectra of GRO~J1655-40 scaled to the case of AGN,
using the Shen et al. (2008) mass distribution. We use the Shen et al. distribution (not the Merloni et al. one) because
it has the mean and FWHM close to those of the Kelly et al. sample.\footnote{The AGN from Kelly et al. (2007)
substitute approximately half of the Kelly et al. (2008) sample of AGN with known masses
and distribution that peaks at $\log({\rm M/M}_{odot})\sim8.8$ and ${\rm FWHM} \sim 1.1$.}
In Figs.~\ref{fig:data}a--b we superimpose the simulated and observed data. It can be seen that the
observations do not overlap with the simulated hard state AGN spectra, while they do occupy the same
region in the plot as the simulated soft state AGN spectra.

\begin{figure}
\centering
\includegraphics[height=5.5cm,bb=165 180 535 580,clip,angle=-90]{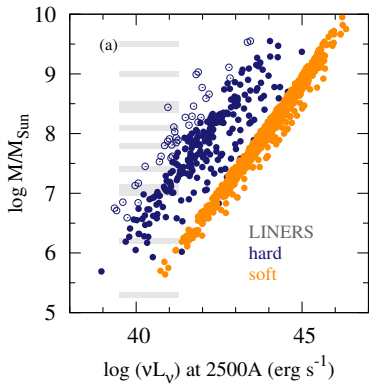}
\caption{Relation between the mass of the black hole in an AGN and the simulated monochromatic
luminosity at 2500\AA\ for mass distribution of LINERS and Type 1 AGN (see
Fig.~\ref{fig:masses}c), hard states - black/blue, soft state - gray/orange, hard state with
insignificant disc component - open circles. The gray rectangles illustrate masses and the
range of $\log (\nu L_{\nu})$ at 2500\AA\/ for the real LINERs from Maoz (2007).}
\label{fig:massluv}
\end{figure}

The interpretation of the spectral states of LINERs (Figs.~\ref{fig:data}c--d) appears more complex since the data points
derived from the real SEDs (Maoz 2007) overlap both with the hard and soft points (simulated based on the LINERs mass
distribution, Fig.~\ref{fig:masses}c). However, it can
be seen in Fig.~\ref{fig:massluv} that in the case of the simulated soft state AGN,
the SEDs that result in $\log(\nu L_\nu)_o \approx 39.5$--41.3 (as reported by Maoz 2007) require
low black hole masses, M/M$_{\odot} < 10^6$, whereas in the case of simulated hard state AGN,
the SEDs require black hole masses of M/M$_{\odot} \sim 10^6$--10$^8$ to reach comparable $\log(\nu L_{\nu})$
at 2500\AA. Since the LINERs in Maoz (2007) have rather high masses (as indicated with the gray
rectangles in Fig.~\ref{fig:massluv}) we conclude that most probably these objects are in a hard spectral state.

In Figs.~\ref{fig:data}e--f we perform similar comparison for a subsample of NLS1 from Grupe et al. (2010).
We use the values of $\alpha_{\rm ox}$ provided by the authors in Table 5, column 6, to calculate
the monochromatic luminosity at 2500\AA\ from tabulated values of the UV slope, $\alpha_{\rm UV}$, and luminosity at
5100\AA\ (columns 5 and 13, respectively). We compare the observations with the simulations for the mass distribution
of the same sample (Fig.~\ref{fig:masses}d). Similarly as in the case of broad line AGN of Kelly et al.,
the NLS1s in the sample of Grupe et al. do not overlap with the hard state simulated SEDs, while they 
do overlap with the simulated AGN in the soft state.

\section{Discussion}
\label{sec:dc}

We studied the scenario in which the main difference between the GBH
and AGN nuclear emission follows from the difference between the masses of the accreting
black holes. We assumed that the geometry of the accretion flow characterized by the ratio
of the heating and cooling compactnesses, $\ell_h/\ell_s$, stays the same for both AGN and
GBHs in a given spectral state. With these assumptions, we used the RXTE archival data of the 
2005 outburst of GRO~J1655-40, a well studied GBH, to simulate an outburst of an AGN hosting
a 10$^9$M$_{\odot}$ black hole. Based on the simulated AGN spectra, we calculated $\alpha_{\rm ox}$,
the simulated X-ray loudness, and $(\nu L_\nu)_o$, the monochromatic luminosity at 2500\AA.
We studied how these two parameters would behave during a hypothetical AGN outburst.

Next, we considered a relation between the simulated $\alpha_{\rm ox}$ and $(\nu L_\nu)_o$ in a
sample of AGN with a range of masses. We simulated the AGN's SEDs for a sample of 1000 AGN with
masses randomly drawn from a log-normal distribution. The mass distributions were observationally
motivated. We considered different classes of AGN: broad line AGN, narrow line Seyfert 1s and LINERs.

Our results suggest that in a sample of AGN with a narrow range of black hole masses one should expect
a deviation from a simple correlation between $\alpha_{\rm ox}$ vs. $(\nu L_\nu)_o$. Instead,
a characteristic U-shape would be displayed. In a sample of AGN with a broad range of black hole masses,
the $\alpha_{\rm ox}$ vs. $(\nu L_\nu)_o$ correlation appears naturally as the result of the mass
spread. The slope of the correlations that we derive from the simulated SEDs depends on the
assumed number of the soft states in the sample, $N_S$, and is consistent with those
observed in real SEDs (e.g Grupe et al. 2010, Lusso et al. 2010, and references therein) for $N_S \geq 70$\%.

We also studied the relation between the simulated $\alpha_{\rm ox}$ and the Eddington luminosity ratio, $L/L_{\rm E}$.
We found that the X-ray loudness correlates with the luminosity in Eddington units down to
$\lambda_{\rm crit} = L_{\rm crit}/L_{\rm E} \approx 0.01$, and the
slope of this correlation roughly matches that reported by Lusso et al. (2010). This result
does not depend either on the adopted AGN mass distribution or on the $N_S$. The correlation
down to $\lambda_{\rm crit}$ is formed by both, soft and hard state simulated SEDs. Below $\lambda_{\rm crit}$
we find only the hard state simulated SEDs, and the correlation changes its sign. This has not been observed 
yet mainly due to the difficulties in determining $\alpha_{\rm ox}$ in low luminosity AGN with optical spectra
often dominated by the host galaxy.
Our result provides thus a prediction for future observational studies.
Interestingly, in a study of X-ray bright optically normal galaxies (XBONGs) Civano et al. (2005) found that the
broad band IR-opt/UV-Xray SEDs of two of their objects can be explained with the Elvis et al. (1994) radio quiet template, that
is characterized by $\alpha_{\rm ox} = 1.38$. The accretion rate of these two XBONGs was estimated at 0.001$L/L_{\rm E}$,
and hence they would fit on the branch of our simulated $\alpha_{ox}$ vs. $L/L_{\rm E}$ correlation with a negative slope,
confirming our prediction that the correlation changes its sign below $\lambda_{\rm crit}$.
Even more intriguingly, the critical
value $\lambda_{\rm crit} \approx 0.01$ matches that for which a change in the sign in the correlation between 
the X-ray photon index, $\Gamma_x$, and Eddington luminosity ratio was reported for AGN (e.g. Green et al. 2009) and
GBHs (Wu \& Gu 2008, Sobolewska et al. in preparation).

Finally, we investigated a possibility of determining the spectral states of AGN by comparing
the simulated data with the observational data of bright Type 1 AGN, NLS1s and
and LINERs. We found that the LINERs are most probably the counterparts of the GBH hard state SEDs,
while the bright Type 1 AGN and NLS1s overlap with the regions occupied by the simulated
soft state SEDs in the $\alpha_{\rm ox}$ vs. $\nu L_{\nu}$ diagram. 

It can be noticed in Fig.~\ref{fig:data}f that the NLS1 luminosities reach $\log(\nu L_{\nu})_o \simeq 46$
while the simulations do not exceed $\log(\nu L_{\nu})_o \simeq 44.5$, even though we use the same sample to determine
the mass distribution for simulations and to compare the simulations with
observations. This difference in luminosity may indicate that the
brightest NLS1s exceed the maximum $L/L_{\rm E}$ in our GRO~J1655-40 template by $\sim1.5$ order of magnitude, which would push the
Eddington ratios of NLS1s to the super Eddington regime of up to 6$L/L_{\rm E}$. Indeed, Grupe et al. (2010) estimated
that in 31 of 43 objects in their NLS1 sample the Eddington ratio exceeds the peak luminosity of GRO~J1655-40 in Eddington units
(17 of them are super Eddington, with the highest estimate of the Eddington ratio being $\sim13L/L_{\rm E}$).

Typical GBHs rarely exceed $0.3L/L_{\rm E}$ (e.g. Done et al. 2007).
The only example of a GBH showing a steady accretion flow at high Eddington fraction is GRS~1915+105
(e.g. Done et al. 2004) that can reach 3 times the Eddington limit.
However, at high $L/L_{\rm E}$ the GBH accretion is most probably influenced by powerful wind and jet outflows
(e.g Janiuk, Czerny \& Siemiginowska 2002, Neilsen \& Lee 2009). Also, at high $L/L_{\rm E}$
the standard Shakura-Sunyaev disc becomes unstable and a different, possibly advection dominated optically thick,
solution may be more appropriate to describe such SEDs (Abramowicz et al. 1988).
They may not be easily scalable between GBHs and AGN and are beyond the scope of the present paper.

Gierli\'nski \& Newton (2006) described two types of the hard-to-soft state transition in GBHs. They differ in terms
of the time needed to change between the states and the Eddington ratios at which the transition is triggered.
The hard-to-soft state transition in the 2005 outburst of GRO~J1655-40 would belong to the dark/fast (DF) category
with $L/L_E \simeq 0.02$ at the beginning of the transition.
That means that our template may lack luminous hard state SEDs. GX~339-4 outburst in 2002/2003 can be an example
of a bright/slow (BS) transition with $L/L_E\simeq 0.2$ when the transition started
(Fender et al. 2004, Gierli\'nski \& Newton 2006, although
notice the uncertainty of GX 339-4 black hole mass estimates, e.g. Hynes et al. 2004, Zdziarski et al. 2004, Munoz-Darias,
Cesares \& Martinez-Pais 2008). The hardness-intensity diagrams for GX 339-4 show that the low state intensity
increases before the transition while the hardness stays approximately constant. It means
that if we included such luminous low state spectra in our template, in the simulated data we would have obtained additional 
points at $\alpha_{ox}\simeq 1$ (constant hardness and hence constant spectral shape) and $\nu L_{\nu}$ at 2500\AA\/ by
up to an order of magnitude higher than at present. Hence, this would not change our conclusions about the spectral
states of NLS1 and Type 1 AGN because these additional hard state simulated points would appear much below the observed
data in both cases (see Fig. \ref{fig:data}a,c).

The analogy between the GBHs and AGN was studied also
by other authors in the context of determining AGN 'states'.
For example, K{\"o}rding et al. (2006) compared the broad band radio-to-X-ray spectra of GBHs and AGN to study
AGN radio states. A number of authors (e.g. Uttley \& McHardy 2005, Ar\'evalo et al. 2006, McHardy et al. 2007)
investigated methods based on time variability. They noticed similarities in the shape of the power density spectra
of GBHs and AGN, which provide an interesting tool in determining the spectral state, but only for a few AGN with good
quality long-term lightcurves.

\subsection{Low/hard state accretion disc in GBHs}

We note that in the hard state the disc of GBHs is cool enough to be shifted away from the RXTE bandpass.
GBHs in the hard state have been observed by modern telescopes with a better low energy coverage than that of RXTE:
Chandra, XMM-Newton, Suzaku and Swift (see e.g. Miller et al. 2002, Reis et al. 2010). These observations revealed a soft
excess reminiscent of an accretion disc with the temperature of $\sim 0.2$ keV extending close to the last
stable orbit. If such soft excess is indeed present in the hard state GBH spectra and if it corresponds to a cool
accretion disc, our parametrization may not be accurate. As a result our values of $\alpha^{\prime}_{\rm GBH}$
parameter in the hard state may be underestimated since the SED at 3 keV will be dominated by the hard X-rays
and not by the disc emission.

GRO~J1655-40 in the hard state was observed by XMM-Newton during the 2005 outburst (Obs ID 0112921301).
This observation was taken $\sim$4.5 hours after the 90428-01-01-03 RXTE observation. We model
these two observations jointly to quantify the effect of the drop in the disc temperature and the shift
of the disc component out of RXTE bandpass.

We apply the spectral model described in Sec.~\ref{sec:data} to the joint RXTE and XMM-Newton data.
We obtain a good agreement of data (0.3--200 keV band) and model, with $\chi^2 = 1089$ for $1081$
degrees of freedom. The temperature of the soft photons converged to kT$_{\rm bb} = 0.166^{+0.011}_{-0.003}$
keV (intermediate between the values reported by Reis et al. 2010 and D{\'{\i}}az Trigo et al. 2007), while we kept
it fixed at 0.4 keV in the case of the RXTE fit. The joint fit resulted in the hard-to-soft compactness
ratio of $\ell_h/\ell_s = 18.0^{+0.8}_{-1.4}$, higher than in the case of modeling only the RXTE data
($\ell_h/\ell_s = 11.8^{+0.6}_{-1.8}$). We were able to put an upper limit on the optical depth of the corona, $\tau < 1.5$
($\tau < 1.0$ for RXTE data only). The Eddington luminosity ratio was $L_{\rm bol}/L_{\rm E} = 7.3 \times 10^{-3}$ 
($6.3 \times 10^{-3}$ calculated based on the best fit model to RXTE data only).

The two models, fitted to the RXTE and RXTE+XMM data, scaled to the case of an AGN with 10$^8$ M$_{\odot}$ gave
$\alpha_{\rm ox} = 0.82$ and 1.06, respectively. Thus, in the hard state spectra it is possible that we underestimate
the simulated $\alpha_{\rm ox}$ by $\sim$25\%, at least in the spectra similar to the one described above, with the
Eddington luminosity ratio lower than 1\%. 
This would however only strengthen the conclusion followed from Fig.~\ref{fig:sim}e
that the $\alpha_{\rm ox}$ vs. Eddington luminosity ratio correlation is expected to change its sign.

Nevertheless, more robust constraints on the temperature of the accretion disc in the hard state GBHs are crucial for the
understanding of the spectral evolution of the accreting black holes.

\section{Conclusions}

\begin{itemize}
 \item We assumed that the main difference between the radiation processes in the AGN and GBHs is induced by the difference in the
black hole mass. We scaled an outburst of a Galactic black hole X-ray binary using an observationally motivated
log-normal AGN mass distribution, and we were able to reconstruct the $\alpha_{\rm ox}$ vs. luminosity at 2500\AA\/ correlation observed
in AGN.
 \item In a sample of AGN with a narrow range of masses we expect a deviation from the  $\alpha_{\rm ox}$ vs. luminosity at
2500\AA\/ correlation.
 \item We predict that the  $\alpha_{\rm ox}$ vs. Eddington luminosity ratio correlation changes its sign for Eddington
accretion rates below 1\%. The XBONGs may be examples of AGN that form the low luminosity branch of the correlation.
 \item Our study suggest that bright Type 1 quasars and NLS1 are in a spectral state similar to the soft state of GBHs,
while LINERS may correspond to the hard state of GBHs.
 \item Tighter constraints on the behaviour of the accretion disc in the hard spectral state in GBHs are needed in order to
understand the balance between the thermal disc emission and the non-thermal Comptonized power-law like hard X-ray emission.
\end{itemize}

\section*{Acknowledgments}
We would like to thank the referee for useful comments and suggestions.
This research was funded in part by NASA contract NAS8-39073.  Partial
support for this work was provided by the {\it Chandra\/} grants
GO8-9125A and GO0-11133X.

\label{lastpage}


\begin{thebibliography}{}

\bibitem[Abramowicz et al.(1988)]{1988ApJ...332..646A} Abramowicz M.~A., 
Czerny B., Lasota J.~P., Szuszkiewicz E., 1988, \apj, 332, 646

\bibitem[Ar{\'e}valo et al.(2006)]{2006MNRAS.372..401A} Ar{\'e}valo P., Papadakis I.~E., Uttley P., McHardy I.~M., 
Brinkmann W., 2006, \mnras, 372, 401 

\bibitem[]{} Arnaud K.~A., 1996, Astronomical Data Analysis Software and Systems V, 101, 17

\bibitem[Avni \& Tananbaum(1982)]{1982ApJ...262L..17A} Avni Y., Tananbaum H., 1982, \apjl, 262, L17 

\bibitem[Bechtold et al.(2003)]{2003ApJ...588..119B} Bechtold J. et al., 2003, \apj, 588, 119 

\bibitem[Belloni et al.(2005)]{2005A&A...440..207B} Belloni T., Homan J., Casella P., van der Klis M.,
Nespoli E., Lewin W.~H.~G., Miller J.~M., M{\'e}ndez M., 2005, \aap, 440, 207

\bibitem[Brocksopp et al.(2006)]{2006MNRAS.365.1203B} Brocksopp C. et al., 2006, \mnras, 365, 1203

\bibitem[Chiang(2002)]{2002ApJ...572...79C} Chiang J., 2002, \apj, 572, 79

\bibitem[Chiang \& Blaes(2003)]{2003ApJ...586...97C} Chiang J., Blaes O., 2003, \apj, 586, 97

\bibitem[Civano et al.(2007)]{2007A&A...476.1223C} Civano F. et al., 2007, \aap, 476, 1223 

\bibitem[]{} Coppi P.S., 1999, in Poutanen J., Svensson R., eds, \aspv 161, High Energy Processes in Accreting Black Holes. \asp, p. 375

\bibitem[Czerny et al.(2009)]{2009ApJ...698..840C} Czerny B., Siemiginowska A., Janiuk A., Nikiel-Wroczy{\'n}ski B.,
Stawarz {\L}., 2009, \apj, 698, 840 

\bibitem[D{\'{\i}}az Trigo et al.(2007)]{2007A&A...462..657D} D{\'{\i}}az Trigo M., Parmar A.~N., Miller J., Kuulkers E.,
Caballero-Garc{\'{\i}}a M.~D., 2007, \aap, 462, 657

\bibitem[Done \& Gierli{\'n}ski(2003)]{2003MNRAS.342.1041D} Done C., Gierli{\'n}ski M., 2003, \mnras, 342, 1041

\bibitem[Done et al.(2004)]{2004MNRAS.349..393D} Done C., Wardzi{\'n}ski G., Gierli{\'n}ski M., 2004, \mnras, 349, 393 

\bibitem[Done et al.(2007)]{2007A&ARv..15....1D} Done C., Gierli{\'n}ski M., Kubota A., 2007, \aapr, 15, 1 

\bibitem[Eracleous et al.(2010)]{2010ApJS..187..135E} Eracleous M., Hwang J.~A., Flohic H.~M.~L.~G., 2010, \apjs, 187, 135 

\bibitem[Fender et al.(2004)]{2004MNRAS.355.1105F} Fender R.~P., Belloni T.~M., Gallo E., 2004, \mnras, 355, 1105

\bibitem[Gierli{\'n}ski et al.(1999)]{1999MNRAS.309..496G} Gierli{\'n}ski M., Zdziarski A.~A., Poutanen J., Coppi P.~S.,
Ebisawa K., Johnson W.~N., 1999, \mnras, 309, 496

\bibitem[Gierli{\'n}ski \& Newton(2006)]{2006MNRAS.370..837G} Gierli{\'n}ski M., Newton J., 2006, \mnras, 370, 837

\bibitem[Goodrich(1989)]{1989ApJ...342..224G} Goodrich R.~W., 1989, \apj, 342, 224

\bibitem[Green et al.(2009)]{2009ApJ...690..644G} Green P.~J. et al., 2009, \apj, 690, 644 

\bibitem[Grupe(2004)]{2004AJ....127.1799G} Grupe D., 2004, \aj, 127, 1799

\bibitem[Grupe et al.(2006)]{2006AJ....131...55G} Grupe D., Mathur S., Wilkes B., Osmer P., 2006, \aj, 131, 55

\bibitem[Grupe et al.(2010)]{2010ApJS..187...64G} Grupe D., Komossa S., Leighly K.~M., Page K.~L., 2010, \apjs, 187, 64 

\bibitem[Gu \& Cao(2009)]{2009MNRAS.399..349G} Gu M., Cao X., 2009, \mnras, 399, 349

\bibitem[Hynes et al.(2004)]{2004ApJ...609..317H} Hynes R.~I., Steeghs D., Casares J.,
Charles P.~A., O'Brien K., 2004, \apj, 609, 317

\bibitem[Janiuk et al.(2002)]{2002ApJ...576..908J} Janiuk A., Czerny B., Siemiginowska A., 2002, \apj, 576, 908 

\bibitem[Kelly et al.(2007)]{2007ApJ...657..116K} Kelly B.~C., Bechtold J., Siemiginowska A.,
Aldcroft T., Sobolewska M., 2007, \apj, 657, 116

\bibitem[Kelly et al.(2008)]{2008ApJS..176..355K} Kelly B.~C., Bechtold J., Trump J.~R., Vestergaard M.,
Siemiginowska A., 2008, \apjs, 176, 355

\bibitem[K{\"o}rding et al.(2006)]{2006MNRAS.372.1366K} K{\"o}rding E.~G., Jester S., Fender R., 2006, \mnras, 372, 1366

\bibitem[Laor(1991)]{1991ApJ...376...90L} Laor A., 1991, \apj, 376, 90

\bibitem[Lusso et al.(2010)]{2010A&A...512A..34L} Lusso E. et al., 2010, \aap, 512, A34

\bibitem[Magdziarz et al.(1998)]{1998MNRAS.301..179M} Magdziarz P., Blaes O.~M., Zdziarski A.~A., Johnson W.~N.,
Smith D.~A., 1998, \mnras, 301, 179

\bibitem[Maoz(2007)]{2007MNRAS.377.1696M} Maoz D., 2007, \mnras, 377, 1696

\bibitem[McHardy et al.(2007)]{2007MNRAS.382..985M} McHardy I.~M., Ar{\'e}valo P., Uttley P., Papadakis I.~E.,
Summons D.~P., Brinkmann W., Page M.~J., 2007, \mnras, 382, 985

\bibitem[McNamara \& Nulsen(2007)]{2007ARA&A..45..117M} McNamara B.~R., Nulsen P.~E.~J., 2007, \araa, 45, 117

\bibitem[Merloni(2003)]{2003MNRAS.341.1051M} Merloni A., 2003, \mnras, 341, 1051

\bibitem[Merloni \& Nayakshin(2006)]{2006MNRAS.372..728M} Merloni A., Nayakshin S., 2006, \mnras, 372, 728

\bibitem[Merloni et al.(2010)]{2010ApJ...708..137M} Merloni A. et al., 2010, \apj, 708, 137

\bibitem[Middleton et al.(2007)]{2007MNRAS.381.1426M} Middleton M., Done C., Gierli{\'n}ski M., 2007, \mnras, 381, 1426

\bibitem[Miller et al.(2002)]{2002MNRAS.335..865M} Miller J.~M., Ballantyne D.~R., Fabian A.~C., Lewin W.~H.~G., 2002, \mnras, 335, 865

\bibitem[Miller et al.(2008)]{2008ApJ...680.1359M} Miller J.~M., Raymond J., Reynolds C.~S., Fabian A.~C.,
Kallman T.~R., Homan J., 2008, \apj, 680, 1359

\bibitem[Mu{\~n}oz-Darias et al.(2008)]{2008MNRAS.385.2205M} Mu{\~n}oz-Darias T., Casares J., Mart{\'{\i}}nez-Pais I.~G.,
2008, \mnras, 385, 2205

\bibitem[Neilsen \& Lee(2009)]{2009Natur.458..481N} Neilsen J., Lee J.~C., 2009, \nat, 458, 481

\bibitem[Osterbrock \& Pogge(1985)]{1985ApJ...297..166O} Osterbrock D.~E., Pogge R.~W., 1985, \apj, 297, 166

\bibitem[Reis et al.(2010)]{2010MNRAS.402..836R} Reis R.~C., Fabian A.~C., Miller J.~M., 2010, \mnras, 402, 836

\bibitem[Remillard \& McClintock(2006)]{2006ARA&A..44...49R} Remillard R.~A., McClintock J.~E., 2006, \araa, 44, 49

\bibitem[Sala et al.(2007)]{2007A&A...461.1049S} Sala G., Greiner J., Vink J., Haberl F., Kendziorra E.,
Zhang X.~L., 2007, \aap, 461, 1049

\bibitem[Shen et al.(2008)]{2008ApJ...680..169S} Shen Y., Greene J.~E., Strauss M.~A., Richards G.~T.,
Schneider D.~P., 2008, \apj, 680, 169

\bibitem[Siemiginowska et al.(2007)]{2007ApJ...657..145S} Siemiginowska A., Stawarz {\L}., Cheung C.~C., Harris D.~E.,
Sikora M., Aldcroft T.~L., Bechtold J., 2007, \apj, 657, 145

\bibitem[Sobolewska et al.(2004)]{2004ApJ...608...80S} Sobolewska M.~A., Siemiginowska A., {\.Z}ycki P.~T., 2004a, \apj, 608, 80

\bibitem[Sobolewska et al.(2004)]{2004ApJ...617..102S} Sobolewska M.~A., Siemiginowska A., {\.Z}ycki P.~T., 2004b, \apj, 617, 102

\bibitem[Sobolewska et al.(2009)]{2009MNRAS.394.1640S} Sobolewska M.~A., Gierli{\'n}ski, M., Siemiginowska A., 2009, \mnras, 394, 1640

\bibitem[Sobolewska \& Papadakis(2009)]{2009MNRAS.399.1597S} Sobolewska M.~A., Papadakis I.~E., 2009, \mnras, 399, 1597

\bibitem[Steffen et al.(2006)]{2006AJ....131.2826S} Steffen A.~T., Strateva I., Brandt W.~N., Alexander D.~M.,
Koekemoer A.~M., Lehmer B.~D., Schneider D.~P., Vignali C., 2006, \aj, 131, 2826

\bibitem[Strateva et al.(2005)]{2005AJ....130..387S} Strateva I.~V., Brandt W.~N., Schneider D.~P., Vanden Berk D.~G.,
Vignali C., 2005, \aj, 130, 387

\bibitem[Uttley \& McHardy(2005)]{2005MNRAS.363..586U} Uttley P., McHardy I.~M., 2005, \mnras, 363, 586

\bibitem[Vasudevan \& Fabian(2009)]{2009MNRAS.392.1124V} Vasudevan R.~V., Fabian A.~C., 2009, \mnras, 392, 1124

\bibitem[Vignali et al.(2003)]{2003AJ....125..433V} Vignali C., Brandt W.~N., Schneider D.~P., 2003, \aj, 125, 433

\bibitem[Wilkes et al.(1994)]{1994ApJS...92...53W} Wilkes B.~J., Tananbaum H., Worrall D.~M., Avni Y., Oey M.~S.,
Flanagan J., 1994, \apjs, 92, 53

\bibitem[Wu \& Gu(2008)]{2008ApJ...682..212W} Wu Q., Gu M., 2008, \apj, 682, 212

\bibitem[Young et al.(2010)]{2010ApJ...708.1388Y} Young M., Elvis M., Risaliti G., 2010, \apj, 708, 1388

\bibitem[Yuan et al.(1998)]{1998A&A...334..498Y} Yuan W., Siebert J., Brinkmann W., 1998, \aap, 334, 498

\bibitem[Zamorani et al.(1981)]{1981ApJ...245..357Z} Zamorani G. et al., 1981, \apj, 245, 357

\bibitem[Zdziarski et al.(2004)]{2004MNRAS.351..791Z} Zdziarski A.~A., Gierli{\'n}ski M., Miko{\l}ajewska J.,
Wardzi{\'n}ski G., Smith D.~M., Harmon B.~A., Kitamoto S., 2004, \mnras, 351, 791

\end{thebibliography}
\end{document}